\documentclass[journal]{IEEEtran}
\usepackage{subfig}
\usepackage{caption}
\usepackage{float}
\usepackage{setspace}
\usepackage{graphicx}
\usepackage{graphicx,dblfloatfix}
\usepackage{tikz}
\usepackage{color}
\usepackage{amsmath}
\usepackage{amssymb} 
\newtheorem{Theorem}{Theorem}
\newtheorem{Lemma}{Lemma}

\usepackage{maplestd2e}
\usetikzlibrary{shapes,arrows,shadows}
\usepackage{xcolor}
\usepackage{lipsum}
\allowdisplaybreaks
\ifCLASSINFOpdf
\else
\fi
\begin{document}
%
\title{Second-Order Perturbation Theory-Based Digital Predistortion for Fiber Nonlinearity Compensation}
%
%
%

\author{O.~S.~Sunish Kumar,~\IEEEmembership{Member,~IEEE,}
	A.~Amari,~\IEEEmembership{Member,~IEEE}, O.~A.~Dobre,~\IEEEmembership{Fellow,~IEEE}, and~R.~Venkatesan,~\IEEEmembership{Life Senior Member,~IEEE}
	
	\thanks{O. S. Sunish Kumar was with the Department of Computer Engineering, Memorial University, St. John’s, Newfoundland and Labrador, A1B 3X5, Canada. He is now with the School of Engineering, Ulster University, Jordanstown, Northern Ireland, BT37 0QB, United Kingdom (e-mail: S.Orappanpara\textunderscore Soman@ulster.ac.uk).}
	
	\thanks{A. Amari is with the VPIphotonics GmbH, Carnotstr. 6, 10587 Berlin, Germany.}
	
	\thanks{O. A. Dobre and R. Venkatesan are with the Department of Computer Engineering, Memorial University, St. John’s, Newfoundland and Labrador, A1B 3X5, Canada.}
	
	\thanks{Manuscript received Month xx, 2021; revised Month xx, 2021.}}

\maketitle

\begin{abstract}
The first-order (FO) perturbation theory-based nonlinearity compensation (PB-NLC) technique has been widely investigated to combat the detrimental effects of the intra-channel Kerr nonlinearity in polarization-multiplexed (Pol-Mux) optical fiber communication systems. However, the NLC performance of the FO-PB-NLC technique is significantly limited in highly nonlinear regimes of the Pol-Mux long-haul optical transmission systems. In this paper, we extend the FO theory to second-order (SO) to improve the NLC performance. This technique is referred to as the SO-PB-NLC. A detailed theoretical analysis is performed to derive the SO perturbative field for a Pol-Mux optical transmission system. Following that, we investigate a few simplifying assumptions to reduce the implementation complexity of the SO-PB-NLC technique. The numerical simulations for a single-channel system show that the SO-PB-NLC technique provides an improved bit-error-rate performance and increases the transmission reach, in comparison with the FO-PB-NLC technique. The complexity analysis demonstrates that the proposed SO-PB-NLC technique has a reduced computational complexity when compared to the digital back-propagation with one step per span.
\end{abstract}


%
\IEEEpeerreviewmaketitle

\section{Introduction}
%
%
%
%
\IEEEPARstart{I}{n} recent years, the core communication network faces a dramatic increase in network traffic. That is fueled by the proliferation of various bandwidth-intensive applications such as virtual reality and cloud services, as well as Internet-of-Things \cite{MChen2019}-\cite{FRestuccia2018}. To accommodate such traffic surges, the optical fibers, such as the standard single-mode fibers (SSMFs), are used in the modern high-speed core communication networks to transmit the broadband information signals over long distances \cite{Infinera2020}. The polarization-multiplexing (Pol-Mux) technology can double the capacity of SSMF by transmitting independent information symbols on the orthogonal polarization tributaries \cite{Amari2017}-\cite{Sunish2018}. However, the electro-optic nonlinearity effect, referred to as the Kerr effect, puts a cap on the maximum achievable transmission reach in the optical communication systems \cite{Amari2017}. 

In single-channel systems, the Kerr-induced signal-signal intra-channel nonlinearity effect is the dominant source of signal distortion that limits the transmission performance \cite{Vassilieva2019}, \cite{Liang2014}. A benchmark technique used for the nonlinearity compensation (NLC) in optical transmission systems is the digital back-propagation (DBP) \cite{JCCartledge2017}. It uses a numerical method to solve the signal propagation equation with appropriate inverted channel parameters \cite{Ip2008}. However, the huge computational load associated with the numerical method limits the practicality of DBP \cite{Ip2008}, \cite{Sunish2017}.

In contrast to the numerical approach, a Volterra series-based analytical method was proposed in \cite{Peddanarappagari1997} to model the nonlinear signal propagation in SSMFs. The results in \cite{Peddanarappagari1997} were later adopted to design a nonlinear equalizer to compensate for the fiber Kerr nonlinearity effect \cite{Guiomar2011}-\cite{Diamantopoulos2019}. However, for long-haul optical links, many Volterra kernels are required in the series expansion to obtain a good approximation of the output optical field. That increases the computational complexity since an order $n$ frequency-domain Volterra kernel entails multiple integrals of order $n$ \cite{Vannucci2002}. 

Alternatively, the solution of the signal propagation equation can be analytically approximated using the first-order (FO) perturbation theory \cite{Vannucci2002}. The NLC technique using this approximate analytical solution is referred to as the FO perturbation theory-based NLC (FO-PB-NLC) \cite{V. Oliari2020}-\cite{Tao2011}. It is interesting to mention that there is an intrinsic relation between
the time-domain perturbation theory and the frequency-domain Volterra series-based approach. For any integer $n$, the order $n$ perturbation solution coincides with order $2n+1$ Volterra series solution \cite{Vannucci2002}. Thus, the perturbation approach is an efficient way to compute the Volterra kernels by avoiding multiple integrations. The main advantages of the PB-NLC techniques are the possibility of a single-stage implementation for the entire fiber link and the symbol rate processing \cite{Tao2011}. However, when the transmit launch power increases, the higher-order perturbation terms become significant, and the compensation performance of the FO-PB-NLC technique decreases \cite{Vannucci2002}.

It is worth noting that the Pol-Mux system has become the norm of transmission in long-haul optical systems. However, such systems are highly vulnerable to the fiber nonlinearity and interference from the co-propagating polarization tributary \cite{Amari2017}. In this paper, we propose a second-order (SO) PB-NLC technique, referred to as SO-PB-NLC, to improve the performance of the FO-PB-NLC technique in a Pol-Mux system. The main contributions of this paper are summarized as follows:
\begin{itemize}
	\item We present a rigorous mathematical analysis to derive the SO nonlinear distortion field expression in a Pol-Mux system with a Gaussian shape assumption for the input pulse shape. Also, we used a symbolic mathematical software Maple for the sanity check of the formulas and to generate corresponding Matlab code for the implementation to avoid any possible error.
	\item We investigate simplifying assumptions to make the expression for the SO nonlinear distortion field less complex. 
	\item We design and implement a digital predistorter using the simplified SO nonlinear distortion field to compensate for the fiber nonlinearity in a Pol-Mux system. 
	\item We carry out a complexity analysis and show that the implementation complexity of the digital predistorter based on the simplified SO distortion field is less when compared to the DBP technique.
	\item We show that the SO-PB-NLC technique provides an extended transmission reach by 15.6\% over the FO-PB-NLC technique, and is only a bit lower in performance when compared with DBP, which has a high implementation complexity.
\end{itemize}

The remainder of this paper is structured as follows. Section II presents the Pol-Mux system model used for the study. Section III explains the theory of the SO-PB-NLC technique for Pol-Mux systems. Numerical simulations and discussions are given in Section IV. Section V concludes the paper. Appendices A and B provide detailed proofs for Lemma 1 and Lemma 2, respectively. Appendix C explains the proof for Theorem 1.     

\textit{Notation}: Lower case italic typeface letters are used for the time-domain representation, whereas the frequency-domain is represented by upper case italic typeface letters. The matrices and vectors are represented by upper case bold typeface letters and lower case bold typeface letters, respectively. 

\section{System Model} 

\subsection{High-Level Description} 

The Pol-Mux system model considered for the study is shown in Fig. 1. 
\begin{figure}[H]
	\centering{}\includegraphics[width=1\columnwidth,height=0.14\paperheight]{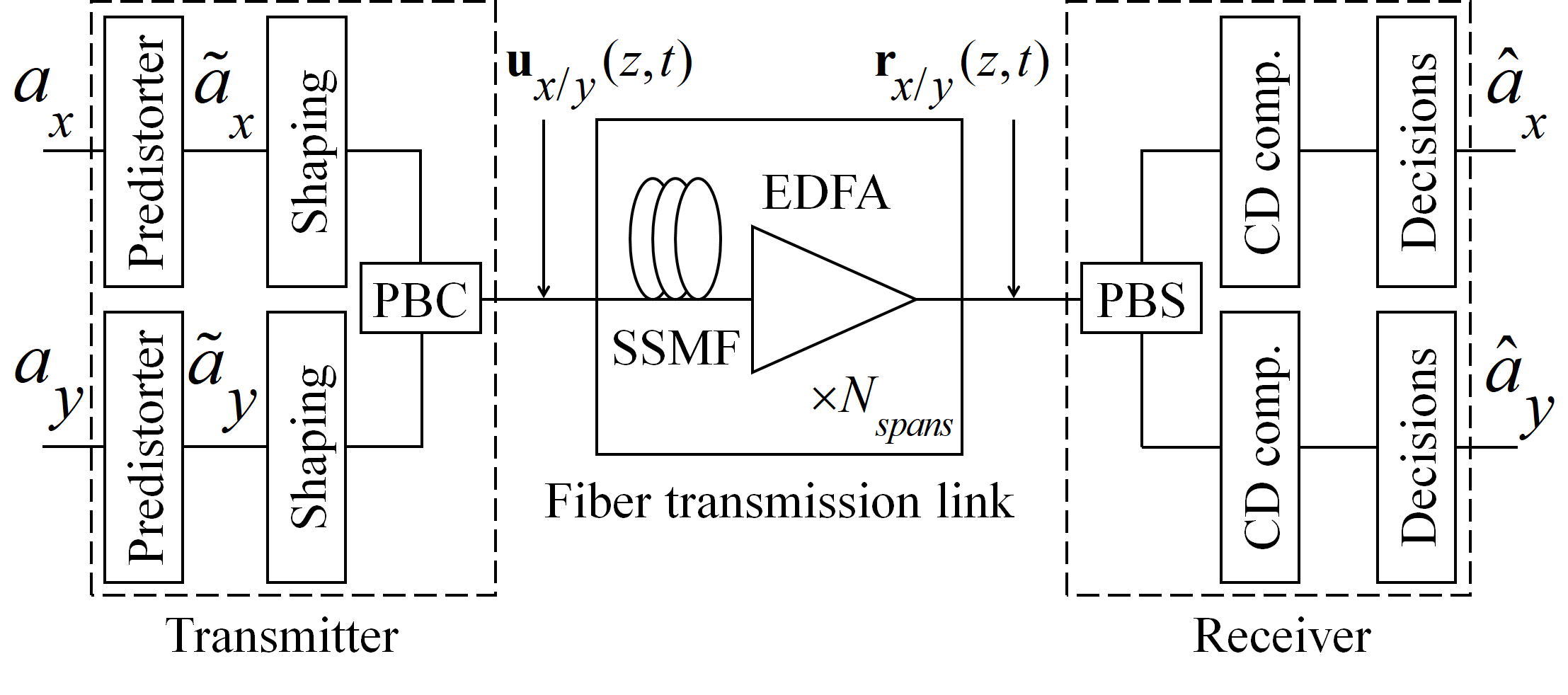}\caption{{The system model for the Pol-Mux system. PBC: polarization beam combiner, SSMF: standard single-mode fiber, EDFA: erbium-doped fiber amplifier, PBS: polarization beam splitter, CD: chromatic dispersion.}}
\end{figure}
In each polarization, a sequence of $K$ symbols $\textbf{a}_{x/y}=\left[a_{1,x/y},a_{2,x/y},\cdots,a_{K,x/y}\right]\in\Omega^{K}$, with $x,$ $y$ as the orthogonal polarization tributaries and $\Omega$ as the symbol alphabet, is processed first to generate the predistorted pulse waveform to combat the nonlinearity effect. Then, the predistorted symbol pulse $\tilde{a}_{x/y}$ is shaped using a pulse shaping filter $\hat{g}(t)$, where $t$ represents the time. The resultant signal in each polarization is represented as $\textbf{u}_{x/y}(z=0,t)=\sum_{\bar{k}=1}^{K}a_{\bar{k},x/y}\hat{g}(t-\bar{k}T),$ with $z$ as the space variable and $T$ as the symbol duration. The pulse shaper is followed by a PBC at the transmitter. The fiber-optic link consists of $N_{spans}$ spans of SSMF and an EDFA in each fiber span to compensate for the fiber attenuation. At the receiver, the signal field $\textbf{r}_{x/y}(z=L,t),$ with $L$ as the transmission length, is demultiplexed using a PBS. The accumulated CD in each polarization is compensated using frequency-domain equalizers. Finally, a symbol-by-symbol maximum likelihood detector is applied, with the resultant estimated symbol represented as $\hat{a}_{x/y}$. 
\subsection{Optical Fiber Channel Model}  

The propagation of the Pol-Mux optical signal through the SSMF can be modeled by using the Manakov equation (noiseless), which is represented as \cite{Vannucci2002}:

\begin{equation}
	\label{eqn1}
	\frac{\partial}{\partial z}{\textrm{\textrm{\textbf{u}}}}+j\frac{\beta_{2}}{2}\frac{\partial^{2}}{\partial t^{2}}{\textrm{\textrm{\textbf{u}}}}=j\frac{8}{9}\gamma({\textrm{\textrm{\textbf{u}}}^{*\dagger}}{\textrm{\textrm{\textbf{u}}}}{\textbf{I}}){\textrm{\textrm{\textbf{u}}}}\exp\left(-\alpha z\right),
\end{equation}
where $\beta_{2}$ is the group velocity dispersion, $\gamma$ is the nonlinearity coefficient, $\alpha$ is the attenuation, $\textbf{\textbf{I}}$ is the identity matrix and the input to the optical fiber is a column vector ${\textrm{\textrm{\textbf{u}}}}(z,t)=[u_{x}(z,t)\,\,\,u_{y}(z,t)]^{\dagger},$ with the superscript $\dagger$ as the transpose. In this study, the split-step Fourier method (SSFM) with a step size of $0.8$ km is used to model the evolution of the modulated optical fields inside the optical fiber channel.

\section{Theory of the SO-PB-NLC Technique for Pol-Mux Systems}  

The Manakov equation governing the evolution of the SO nonlinear distortion field can be represented as \cite{Tao2011}:
\begin{multline}
	\label{eqn2}
	\frac{\partial}{\partial z}u_{2,x/y}(z,t)=\underset{\textrm{Linear part}}{\underbrace{-j\frac{\beta_{2}}{2}\frac{\partial^{2}}{\partial t^{2}}u_{2,x/y}(z,t)}}+\exp(-\alpha z)\\
	\underset{\textrm{Nonlinear part}}{\underbrace{\left(\begin{array}{c}
				j\underset{\textrm{Term 1}}{\underbrace{2\left(\left|u_{0,x/y}(z,t)\right|^{2}+\left|u_{0,y/x}(z,t)\right|^{2}\right)\tilde{u}_{1,x/y}(z,t)}}\\
				+j\underset{\textrm{Term 2}}{\underbrace{\left(u_{0,x/y}^{2}(z,t)+u_{0,y/x}^{2}(z,t)\right)\tilde{u}_{1,x/y}^{*}(z,t)}}
			\end{array}\right)}},
\end{multline}
where $\tilde{u}_{1}$ is the FO field distorted by CD while evolving through the optical fiber link. The dispersed FO perturbative field (or ghost pulse) is used in the calculation of the SO perturbative field. The linear part in (\ref{eqn2}) represents the dispersion of the SO distortion field $u_{2,x/y}$ when it evolves through the optical fiber link. The nonlinear part has two terms: Term 1, which represents the intra-channel cross-phase modulation between the zeroth-order (or linearly dispersed) pulse and the dispersed FO ghost pulse; and Term 2, which is the intra-channel four-wave mixing (FWM) term. \\
The modulated input pulse sequence can be represented as: 
\begin{equation}
\label{eqn3}
u_{0}(z=0,t)=\sqrt{P_{0}}\sum_{\bar{k}}a_{\bar{k}}\hat{g}(z=0,t-\bar{k}T),
\end{equation}
where $a_{\bar{k}}$ is the data information of the $\bar{k}^{th}$ pulse, $P_{0}$ is the pulse peak power, and $\hat{g}(z,t)$ is the pulse temporal waveform at $z.$ By substituting (\ref{eqn3}) in Term 1 of (\ref{eqn2}), we get the expansion for Term 1 as:
\begin{multline}
\label{eqn4}
P_{0}^{5/2}\sum_{m}\sum_{n}\sum_{l}\sum_{k}\sum_{p}
\left(a_{m,x/y}a_{l,x/y}^{*}+a_{m,y/x}a_{l,y/x}^{*}\right)\\\times a_{n,x/y}
\left(a_{k,x/y}a_{p,x/y}^{*}+a_{k,y/x}a_{p,y/x}^{*}\right)\\
\times\tilde{g}_{1,m+n-l}(z,t-
(m+n-l)T)\hat{g}_{k}(z,t-kT)\hat{g}_{p}^{*}(z,t-pT),
\end{multline}   
where $m, n, l, k, $ and $p$ represent the pulse indices, and their values vary between $-\frac{L_{w}}{2}$ to $+\frac{L_{w}}{2}$, with $L_{w}$ as the symbol window length to calculate the FO/SO nonlinear distortion fields. Similarly, by substituting (\ref{eqn3}) in Term 2 of (\ref{eqn2}), we get the expansion for Term 2 as:
\begin{multline}
\label{eqn5}
P_{0}^{5/2}\sum_{m}\sum_{n}\sum_{l}\sum_{k}\sum_{p}
\left(a_{m,x/y}^{*}a_{l,x/y}+a_{m,y/x}^{*}a_{l,y/x}\right)\\ \times a_{n,x/y}^{*}\left(\vphantom{a_{m,x/y}^{*}}a_{k,x/y}a_{p,x/y}+a_{k,y/x}a_{p,y/x}\right)\\
\times\tilde{g}_{1,m+n-l}^{*}(z,t-(m+n-l)T)
\hat{g}_{k}(z,t-kT)\hat{g}_{p}(z,t-pT).
\end{multline}
From (\ref{eqn4}) and (\ref{eqn5}), it is clear that the SO nonlinear distortion field is induced by the nonlinear beating between quintuplet pulses. The schematic diagram shown in Fig. 2 illustrates the quintuplet interaction to generate the SO nonlinear distortion field in (\ref{eqn4}) (i.e., Term 1 of (\ref{eqn2})). 
\begin{figure}[h]
	\centering{}\includegraphics[width=1\columnwidth,height=0.09\paperheight]{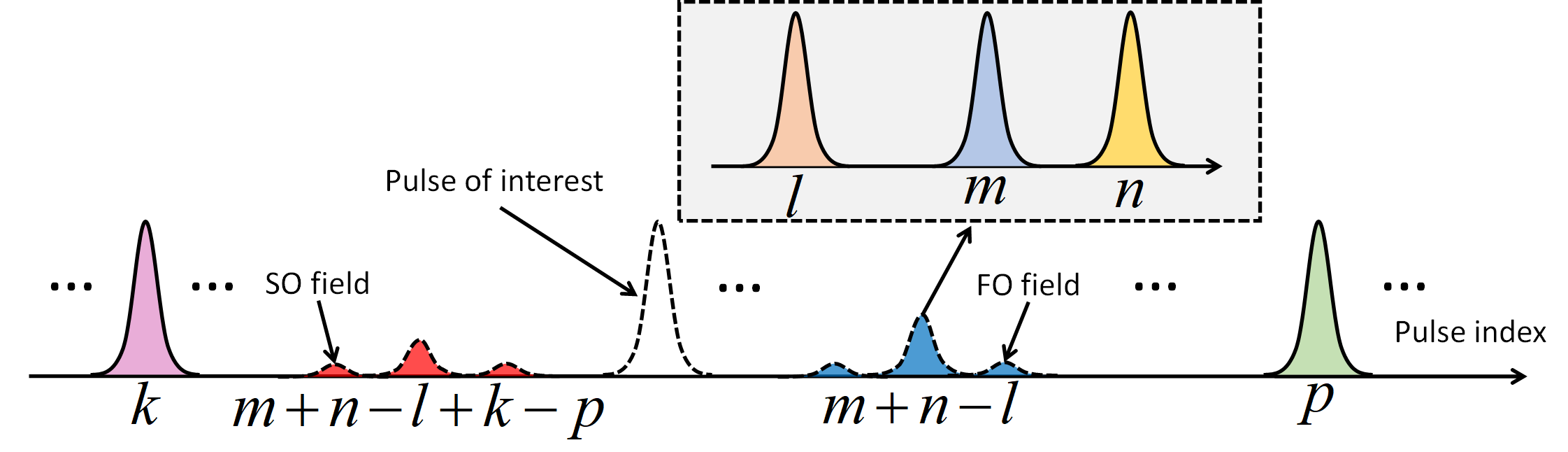}\caption{{Illustration showing the quintuplet pulse interaction to generate the SO nonlinear distortion field.}}
\end{figure}

In the FO perturbation theory, the triplet pulses located at time indices $m, n,$ and $l$ (please see the inset in Fig. 2) induce a FO ghost pulse at the time index $m+n-l$ according to the FWM theory \cite{Tao2011}-\cite{Vassilieva2018}. In congruous with the FO perturbation theory and based on the FWM theory, the SO ghost pulse is generated at the time index $m+n-l+k-p$ by the nonlinear beating between two zeroth-order pulses located at time indices $k$ and $p$ and the FO ghost pulse located at the time index $m+n-l$. For Term 1, the phase-matching condition to induce the SO ghost pulse at the pulse of interest (assuming the pulse of interest is located at the zeroth time index) is $m+n-l+k-p=0$ or $p=m+n-l+k$. Similarly, for Term 2, the phase-matching condition is $p=m+n-l-k$.

Fig. 3 shows a detailed illustration of the SO nonlinear distortion field generation by the nonlinear interaction between quintuplet pulses. The middle sub-figure shows the pulse overlap caused by the CD-induced pulse spreading. These dispersed quintuplet pulses interact nonlinearly during its propagation along the fiber to generate the SO nonlinear distortion field. The bottom sub-figure indicates how the FO and SO nonlinear distortion fields are added to the pulse of interest. 
\begin{figure}[h]
	\centering{}\includegraphics[width=1\columnwidth,height=0.18\paperheight]{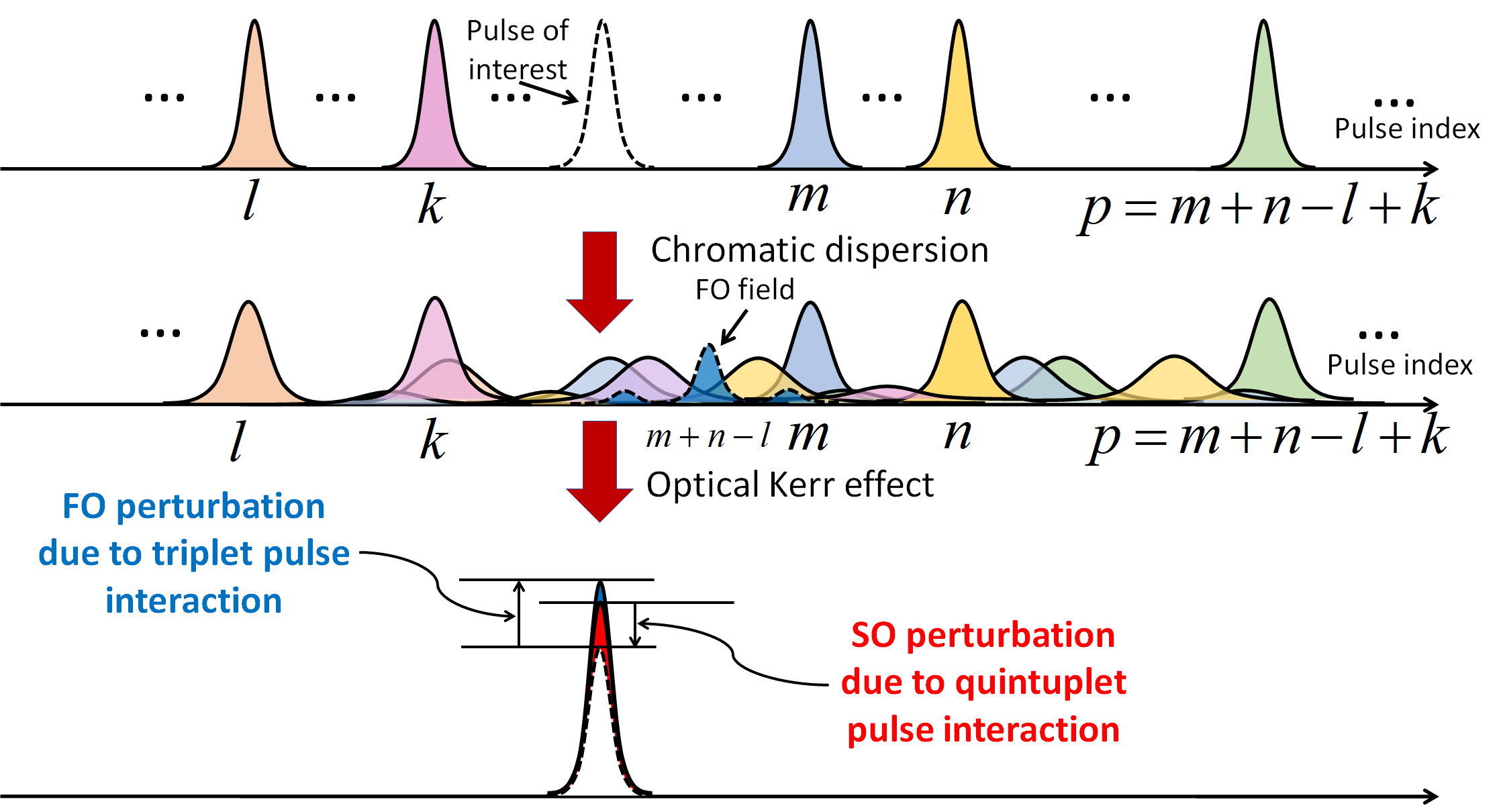}\caption{{The quintuplet pulses involved in the SO distortion field calculation, adapted from \cite{Tao2011}-\cite{Vassilieva2018}.}}
\end{figure}

It can be observed from Fig. 3 that the nonlinear interaction between the FO ghost pulse at an arbitrary time index $m+n-l$ and the linearly dispersed pulses at the time indices $k$ and $p$ lead to a three-pulse collision scenario, and its impact is negligible because of the reduced pulse overlap between the pulse of interest and the other interacting pulses \cite{DarFeder2016}. In other words, the pulse interaction is prevalent only for the FO ghost pulse induced at the symbol under consideration and the dispersed pulses at the time indices $k$ and $p$. To reduce the complexity of the SO-PB-NLC technique, we neglect the FO fields generated at the time indices other than that of the pulse at the zeroth index for which the SO distortion field is calculated. That can be achieved by substituting the phase-matching condition $l=m+n$ in (\ref{eqn4}) and (\ref{eqn5}). That leads to the SO phase-matching conditions $p=k$ for Term 1 and $p=-k$ for Term 2. Figs. 4 and 5 show the nonlinear interaction between the quintuplet pulses in Term 1 and Term 2 of (\ref{eqn2}), respectively.  
\begin{figure}[h]
	\begin{centering}
		\includegraphics[width=1\columnwidth,height=0.18\paperheight]{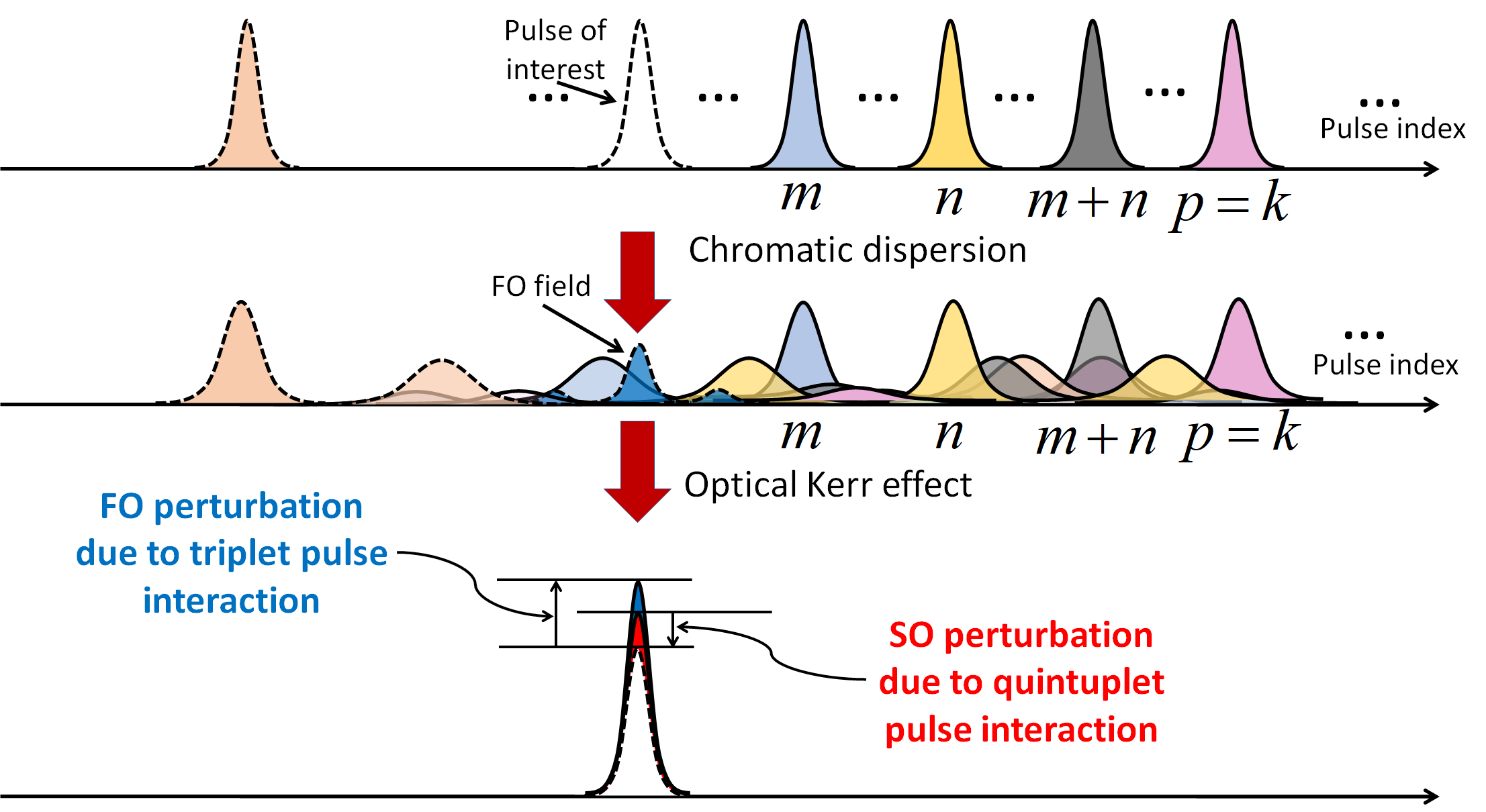}
		\par\end{centering}
	\centering{}\caption{The quintuplet pulses involved in Term 1 of (\ref{eqn2}) and their nonlinear interaction for the phase-matching conditions $l=m+n$ and $p=k$, adapted from \cite{Tao2011}-\cite{Vassilieva2018}.}
\end{figure}

In Fig. 4, the phase-matching condition $l=m+n$ leads to $p=k$. The middle sub-figure indicates the nonlinear interaction between the dispersed FO ghost pulse at the zeroth time index and the other linearly dispersed pulse. That results in a two-pulse collision scenario, as given in \cite{DarFeder2016} and \cite{Vassilieva2018}, and contributes more to the SO ghost pulse generation at the pulse of interest. The bottom sub-figure shows the corresponding SO field generation. On the other hand, in Fig. 5, the phase-matching condition $l=m+n$ leads to $p=-k$, which increases the chance of constructive/destructive interference caused by the three-pulse collision between the FO ghost pulse at the zeroth time index and two linearly dispersed pulses at $p=-k$ and $k$, as shown in the middle sub-figure \cite{DarFeder2016}, \cite{DarFeder2015}. The bottom sub-figure depicts the corresponding SO distortion field added to the pulse of interest.  

\begin{figure}[H]
	\begin{centering}
		\includegraphics[width=1\columnwidth,height=0.18\paperheight]{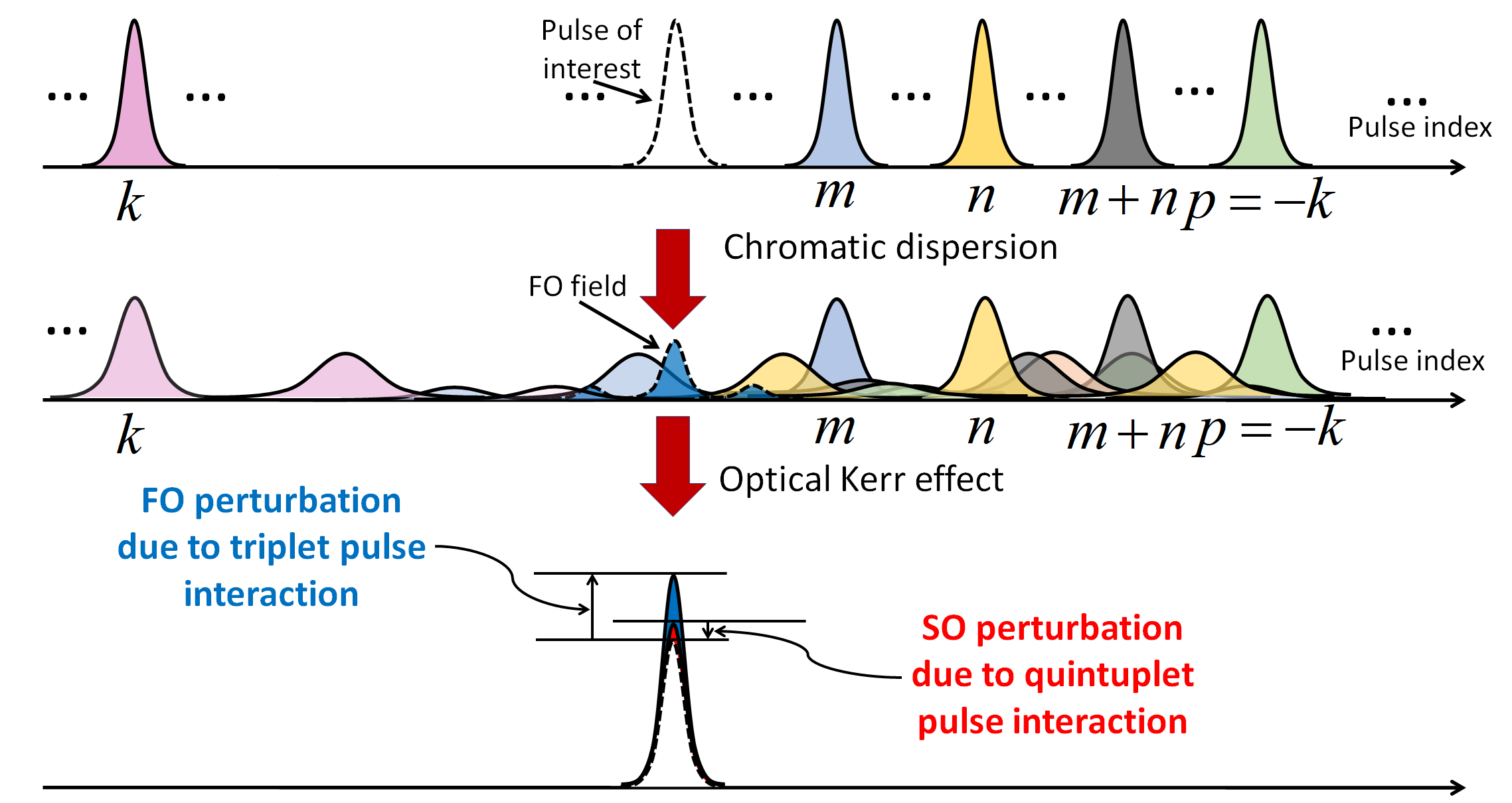}
		\par\end{centering}
	\centering{}\caption{The quintuplet pulses involved in Term 2 of (\ref{eqn2}) and their nonlinear interaction for the phase-matching conditions $l=m+n$ and $p=-k$, adapted from \cite{Tao2011}-\cite{Vassilieva2018}.}
\end{figure}

For simplicity of presentation, Term 1 and Term 2 of the nonlinear part in (\ref{eqn2}) are first considered separately, and then, combined.     

\begin{Lemma} 
	\textit{By considering Term 1 of the nonlinear part in (\ref{eqn2}), the coefficient of the nonlinear interaction between five input Gaussian pulses $\sqrt{P_{0}}a_{m/n/l/k/p}\exp(-(t-T_{m/n/l/k/p})^{2}/2\tau^{2})$, where $\tau$ is the pulse width, at five time indices $T_{m},\,T_{n},\,T_{l},\,T_{k},\,T_{p}$ with the assumption of a symbol rate operation (i.e., $t=0$) and substituting the phase-matching conditions $l=m+n$ and $p=k$, can be expressed as (\ref{eqn6}).}
\end{Lemma}	
	\textit{Proof}: Please refer to Appendix A.

It is important to mention that the Gaussian shape assumption for the input pulse shape is adopted to simplify the mathematical analysis, as in \cite{Tao2011}. In the perturbative analysis, the nonlinear perturbation coefficients are calculated using the overlap integrals that relate the symbol under consideration to other symbols that take part in the nonlinear interaction. The overlap integrals cannot be calculated analytically for non-Gaussian pulse shapes, such as root-raised cosine (RRC) or Nyquist pulses. In \cite{GhazisaeidiEssiambre2014}, a stationary-phase approximation is adopted to circumvent the difficulty in explicitly evaluating overlap integrals for Nyquist pulses. The results in \cite{OyamaNakashima2013} indicate that using an optimized scaling factor, the Gaussian pulse shape assumption in the perturbative analysis is reasonably valid for systems using RRC pulse shape to demonstrate the proof of concept. It is worth mentioning that in the SO-PB-NLC technique, the large CD assumption in \cite{Tao2011} (i.e., $\beta_{2}z\gg \tau^{2}$) is relaxed, such that the SO-PB-NLC technique can reduce the approximation error and improve the NLC performance gain. 

\begin{table*}[!t]
	\setlength{\jot}{-8pt}
	\begin{multline}
		\label{eqn6}
		\widetilde{\textbf{C}}_{m,n,k}^{\text{SO,\,Term\,1}}=-\tau^{4}\intop_{0}^{L}\intop_{0}^{z}\frac{\exp(-\alpha(z+s))}{\sqrt{-\widetilde{A}(z,s)\widetilde{B}(s)}}\exp\left\{ \frac{T^{2}}{\widetilde{A}(z,s)\widetilde{B}(s)}\right.\left[\check{A}_{m,n,k}\tau^{6}\right.
		+2j\beta_{2}(\check{B}_{m,n,k}z\\
		+\check{C}_{m,n,k}s)\tau^{4}+3\beta_{2}^{2}(\check{D}_{m,n,k}z^{2}-\check{E}_{m,n,k}sz+k^{2}s^{2})\tau^{2}
		\left.\left.-5jmnsz^{2}\beta_{2}^{3}\right]\vphantom{\frac{T^{2}}{\check{A}(z,s)\overline{C}(s)}}\right\} \,ds\,dz,
	\end{multline}
	\begin{raggedright}
		where
		\par\end{raggedright}
	\centering{}%
	\setlength{\jot}{0pt}
	\begin{minipage}[t]{0.382\paperwidth}%
		\begin{flalign}
			& \widetilde{A}(z,s)=(j\tau^{6}-3\beta_{2}\left(s+2/3z\right)\tau^{4}\nonumber \\
			& \hspace{1.3cm}-6j\beta_{2}^{2}\left(s-7/6z\right)z\tau^{2}-5sz^{2}\beta_{2}^{3}),\\
			& \widetilde{B}(s)=(j\tau^{2}+\beta_{2}s),\\
			& \check{A}_{m,n,k}=({\displaystyle k^{2}+m^{2}+nm+n^{2}}),
		\end{flalign}
	\end{minipage}%
	\begin{minipage}[t]{0.382\paperwidth}%
		\begin{flalign}
			& \check{B}_{m,n,k}=({\displaystyle m^{2}+\left(-2k+n\right)m}\nonumber \\
			& \hspace{4cm}-2kn+n^{2}),\\
			& \check{C}_{m,n,k}={\displaystyle ({\displaystyle k^{2}-3/2nm}),}\\
			& {\displaystyle \check{D}_{m,n,k}=({\displaystyle m^{2}-1/3nm+n^{2}}),}\\
			& \check{E}_{m,n,k}=({\displaystyle 4/3((k-3/2n)m+kn)}).
		\end{flalign}
	\end{minipage}
	\rule[0.5ex]{1\textwidth}{0.5pt}
\end{table*}

\begin{Lemma} 
	\textit{By considering Term 2 of the nonlinear part in (\ref{eqn2}), the coefficient of the nonlinear interaction between five input Gaussian pulses $\sqrt{P_{0}}a_{m/n/l/k/p}\exp(-(t-T_{m/n/l/k/p})^{2}/2\tau^{2})$ at five time indices $T_{m},\,T_{n},\,T_{l},\,T_{k},\,T_{p}$ with the assumption of a symbol rate operation and substituting the phase-matching conditions $l=m+n$ and $p=-k$, can be expressed as (\ref{eqn14}).}
\end{Lemma}
	\textit{Proof}: Please refer to Appendix B.
\begin{table*}[!t]
	\setlength{\jot}{-8pt}
	\begin{multline}
		\label{eqn14}
		\widetilde{\textbf{C}}_{m,n,k}^{\text{SO,\,Term\,2}}=\sqrt{3}\tau^{4}\intop_{0}^{L}\intop_{0}^{z}\frac{\sqrt{\widehat{A}(z,s)}\exp(-\alpha(z+s))}{\sqrt{\widehat{B}(z,s)\widehat{C}(z)}\left(\sqrt{-\widetilde{B}(s)\widehat{D}(z,s)}\right)^{*}}\exp\left\{ \frac{-jT^{2}}{\tau^{2}\widehat{B}(z,s)\widehat{E}(s)}\right.\left[\check{A}_{m,n,k}\tau^{6}\right.\\
		-j\beta_{2}(\breve{A}_{m,n,k}z+\breve{B}_{m,n,k}s)\tau^{4}+\beta_{2}^{2}s(\breve{C}_{m,n,k}z+3k^{2}s)\tau^{2}\left.\left.+2jk^{2}s^{2}z\beta_{2}^{3}\right]\vphantom{\vphantom{\frac{-jT^{2}}{\tau^{2}\widehat{B}(z,s)\widehat{F}(s)}}}\right\} \,ds\,dz,
	\end{multline}
	\begin{raggedright}
		where
		\par\end{raggedright}
	\begin{centering}
		\setlength{\jot}{0pt}
		\begin{minipage}[t]{0.382\paperwidth}%
			\begin{flalign}
				& \widehat{A}(z,s)=({\displaystyle {\it j\tau^{4}+js}z\beta_{2}^{2}+3\tau^{2}\beta_{2}(s-z)}),\\
				& {\displaystyle \widehat{B}(z,s)=({\displaystyle \tau^{4}-3j(s-7/3z)\beta_{2}\tau^{2}}}\nonumber \\
				& \hspace{5cm}+5sz\beta_{2}^{2}),\\
				& {\displaystyle \widehat{C}(z)=({\displaystyle j\tau^{2}+\beta_{2}z}),}\\
				& \widehat{D}(z,s)={\displaystyle ({\displaystyle \tau^{4}+sz\beta_{2}^{2}+{\it j3}\tau^{2}\beta_{2}(s-z)}),}
			\end{flalign}
		\end{minipage}%
		\begin{minipage}[t]{0.382\paperwidth}%
			\begin{flalign}
				& \widehat{E}(s)=({\displaystyle j\tau^{2}-\beta_{2}s}),\\
				& {\displaystyle \breve{A}_{m,n,k}=({\displaystyle -6k^{2}+4(m+n)k}}\nonumber \\
				& \hspace{3cm}-3m^{2}+nm-3n^{2}),\\
				& \breve{B}_{m,n,k}={\displaystyle ({\displaystyle 2k^{2}-3nm}),}\\
				& {\displaystyle \breve{C}_{m,n,k}=({\displaystyle {\displaystyle -4k^{2}+4k(m+n)-5nm}}).}
			\end{flalign}
		\end{minipage}
		\par\end{centering}
	\rule[0.5ex]{1\textwidth}{0.5pt}
\end{table*}
\begin{Theorem} 
	\textit{For the case of the transmission of a Pol-Mux optical signal through the SSMF, the five input Gaussian pulses $\sqrt{P_{0}}a_{m/n/l/k/p,x/y}\exp(-(t-T_{m/n/l/k/p})^{2}/2\tau^{2})$ at five time indices $T_{m},\,T_{n},\,T_{l},\,T_{k},\,$ and $T_{p}$, with the phase matching conditions $l=m+n$ and $p=k$ for Term 1 and $l=m+n$ and $p=-k$ for Term 2, generate the SO ghost pulse at the zeroth time index; with the assumption of a symbol rate operation, this can be expressed as:} 
\end{Theorem}
\begin{multline}
	\label{eqn23}
	\widetilde{u}_{2,x/y}(L,t)=\frac{64}{81}\gamma^{2}\varepsilon_{SO}P_{0}^{5/2}\sum_{m}\sum_{n}\sum_{k}\\
	\left[\vphantom{\widetilde{\textbf{C}}_{m,n,k}^{SO,\,Part\,2}}2\left(a_{m,x/y}a_{m+n,x/y}^{*}+a_{m,y/x}a_{m+n,y/x}^{*}\right)a_{n,x/y}\right.\\
	\times\left(a_{k,x/y}a_{k,x/y}^{*}+a_{k,y/x}a_{k,y/x}^{*}\right)\widetilde{\textbf{C}}_{m,n,k}^{\text{SO,\,Term\,1}}\\+\left(a_{m,x/y}^{*}a_{m+n,x/y}+a_{m,y/x}^{*}a_{m+n,y/x}\right)\\
	\left.\times a_{n,x/y}^{*}\left(a_{k,x/y}a_{-k,x/y}+a_{k,y/x}a_{-k,y/x}\right)\widetilde{\textbf{C}}_{m,n,k}^{\text{SO,\,Term\,2}}\right],
\end{multline}
\textit{where}  $\varepsilon_{SO}$ \textit{is a scaling factor account for the Gaussian pulse shaping assumption and the power uncertainty in the fiber span and} $\widetilde{\textbf{C}}_{m,n,k}^{\text{SO,\,Term\,1}}$ \textit{and} $\widetilde{\textbf{C}}_{m,n,k}^{\text{SO,\,Term\,2}}$ \textit{are given by (\ref{eqn6}) and (\ref{eqn14}), respectively.} \\
\textit{Proof}: Please refer to Appendix C.

\begin{figure}[h]
	\begin{centering}
		\includegraphics[width=1\columnwidth,height=0.18\paperheight]{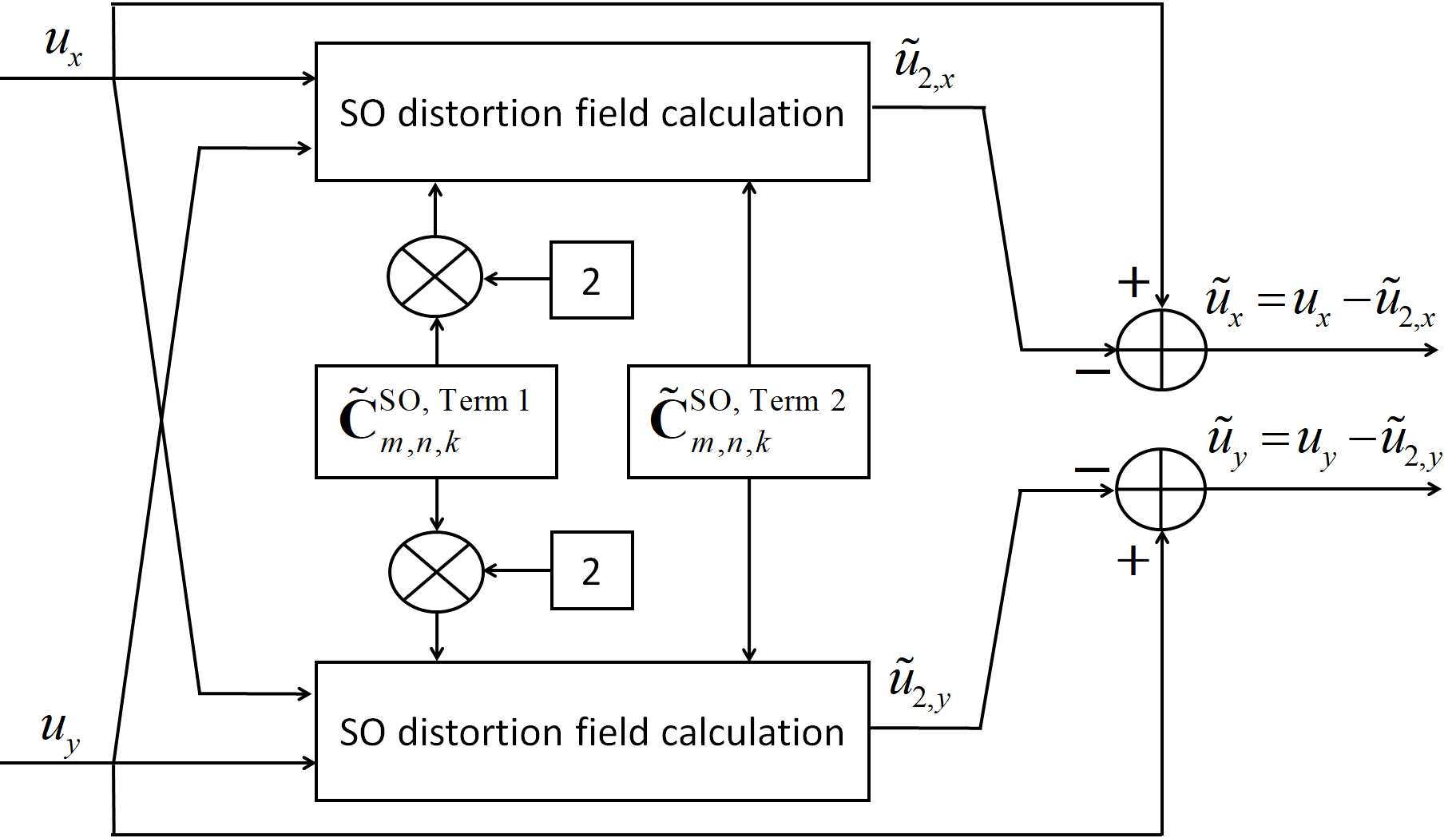}
		\par\end{centering}
	\centering{}\caption{Block diagram of SO-PB-NLC technique using (\ref{eqn23}).}
\end{figure} 
It is worth mentioning that the nonlinearity coefficients $\widetilde{\textbf{C}}_{m,n,k}^{\text{SO,\,Term\,1}}$ and $\widetilde{\textbf{C}}_{m,n,k}^{\text{SO,\,Term\,2}}$ are same as those derived for single-polarization systems. However, for the Pol-Mux system, the SO nonlinear distortion field in (\ref{eqn23}) has an additional cross-polarization interference (cross-talk) term and a scaling factor $\frac{8}{9}$ for $\gamma$  when compared to the single-polarization case. Fig. 6 shows the block diagram of the SO-PB-NLC technique using (\ref{eqn23}) for Pol-Mux systems. The 3-D nonlinearity coefficient matrices $\widetilde{\textbf{C}}_{m,n,k}^{\text{SO,\,Term\,1}}$ and $\widetilde{\textbf{C}}_{m,n,k}^{\text{SO,\,Term\,2}}$ are calculated offline and stored in look-up tables (LUTs). Then, the nonlinear distortion field is calculated using (\ref{eqn23}), which is followed by the subtraction of the calculated field from the input field to generate the predistorted signal. In the proposed SO-PB-NLC technique, the nonlinearity coefficients in $\widetilde{\textbf{C}}_{m,n,k}^{\text{SO,\,Term\,1}}$ and $\widetilde{\textbf{C}}_{m,n,k}^{\text{SO,\,Term\,2}}$ are truncated by a threshold value given as $20\textrm{\,log}_{10}\left(\left|\widetilde{\textbf{C}}_{m,n,k}^{\text{SO,\,Term\,1/Term\,2}}\right|/\left|\widetilde{C}_{0,0,0}^{\text{SO,\,Term\,1/Term\,2}}\right|\right)<\mu$ dB \cite{Tao2011}. 
\begin{figure}[h]
	\centering{}\includegraphics[width=1\columnwidth,height=0.17\paperheight]{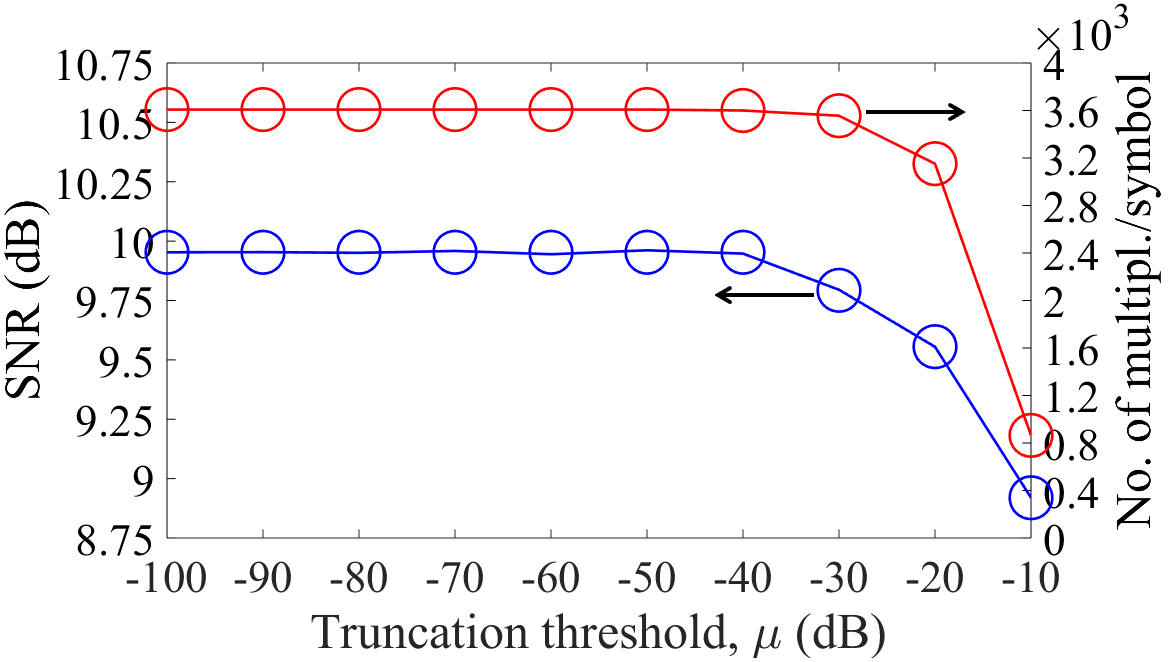}\caption{The impact of the SO coefficient truncation threshold $\mu$ on the performance of the SO-PB-NLC technique.}
\end{figure}
In this work, a coefficient truncation threshold of $\mu=-40$ dB is used to reduce the effective number of the perturbation terms to reduce the implementation complexity. The selection of the truncation threshold $\mu=-40$ dB is based on the results in Fig. 7. 

In implementing the SO-PB-NLC technique, we consider a symbol window size of $L_{w}=100$ to calculate the SO nonlinear distortion field added to the symbol under consideration. For the first/last symbol in a given frame, the magnitude of the symbols outside the selected window to the left/right of the symbol under consideration is considered as zero. It is important to mention that this assumption does not affect the accuracy of the bit-error-rate (BER) calculation since we neglect several symbols from the beginning and the end of a given frame to account for the filter-induced delays in the transmission link. In Fig. 7, it is observed that the signal-to-noise ratio (SNR) improvement for the SO-PB-NLC technique with a coefficient truncation threshold higher than $\mu=-40$ dB is negligibly small. Based on this observation, we select a truncation threshold $\mu=-40$ dB in the implementation of the SO-PB-NLC technique with a symbol window size of $L_{w}=100.$ Fig. 7 also shows the complexity in terms of the number of real-valued multiplications/symbol. It is also observed from Fig. 7 that the increase in the number of perturbation terms beyond the truncation threshold of $\mu=-40$ dB is negligibly small for a selected symbol window size of $L_{w}=100.$

\begin{figure}[!t]
	\centering{}\includegraphics[width=1\columnwidth,height=0.17\paperheight]{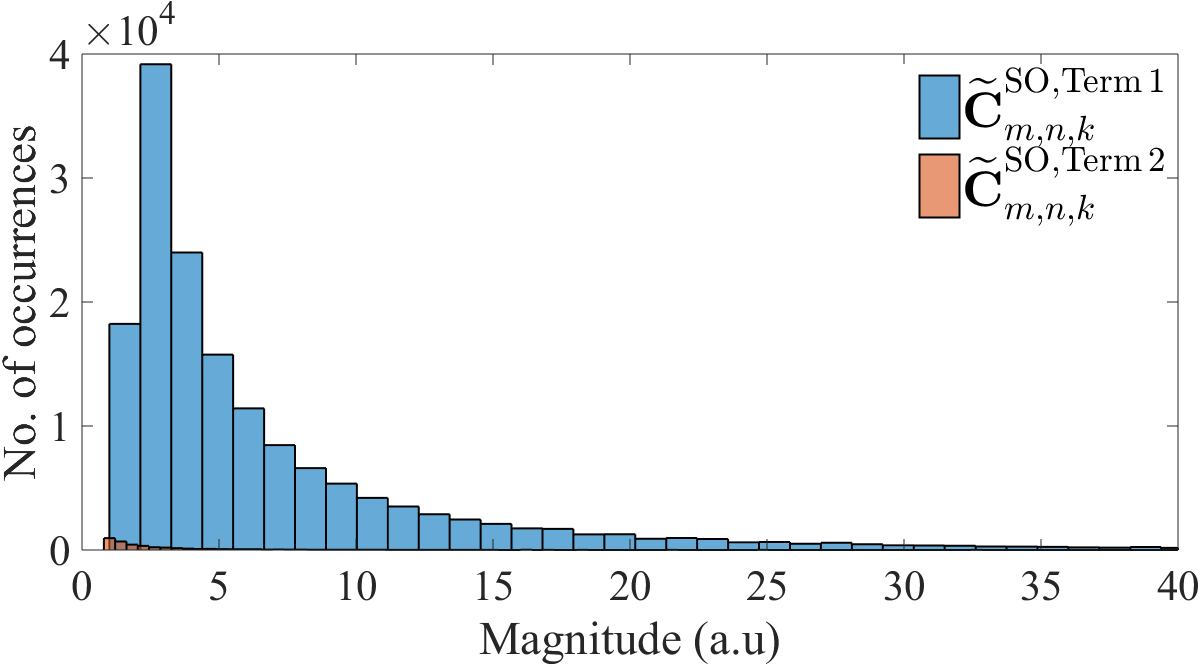}\caption{{The number of occurrences of the magnitude of the nonlinearity coefficients in $\widetilde{\textbf{C}}_{m,n,k}^{\text{SO,\,Term\,1}}$ and $\widetilde{\textbf{C}}_{m,n,k}^{\text{SO,\,Term\,2}}.$}}
\end{figure}

To count the frequency of occurrence of nonlinearity coefficients in $\widetilde{\textbf{C}}_{m,n,k}^{\text{SO,\,Term\,1}}$ and $\widetilde{\textbf{C}}_{m,n,k}^{\text{SO,\,Term\,2}}$ that satisfy the threshold condition, we plot the histogram of the coefficient magnitudes in Fig. 8. It is interesting to note that the number of occurrence of coefficients in $\widetilde{\textbf{C}}_{m,n,k}^{\text{SO,\,Term\,2}}$ above the truncation threshold is significantly lower than that in $\widetilde{\textbf{C}}_{m,n,k}^{\text{SO,\,Term\,1}}$. Also, the magnitudes of the coefficients are close to zero, with a significantly lower variance. That may be due to the constructive/destructive interference caused by the three-pulse collision between the FO ghost pulse and the linearly dispersed pulses \cite{DarFeder2016}, \cite{DarFeder2015}. We investigate the performance of the SO-PB-NLC technique with and without considering Term 2 of (\ref{eqn2}) through numerical simulations in Section IV. 

In the implementation of the SO-PB-NLC technique, we adopt a quantization and combination method proposed in \cite{ZTao2013} to reduce the computational complexity further. It is important to mention that the nonlinearity coefficients in $\widetilde{\textbf{C}}_{m,n,k}^{\text{SO,\,Term\,1}}$ and $\widetilde{\textbf{C}}_{m,n,k}^{\text{SO,\,Term\,2}}$ are very similar, in particular for those with large indexes. Based on this fact, we round the real and imaginary parts of the nonlinearity coefficients to the nearest integer. The FO/SO coefficient quantization and the triplet/quintuplet combination dramatically reduce the computational complexity. 
\begin{figure}[t]
	\begin{centering}
		\includegraphics[width=1\columnwidth,height=0.14\paperheight]{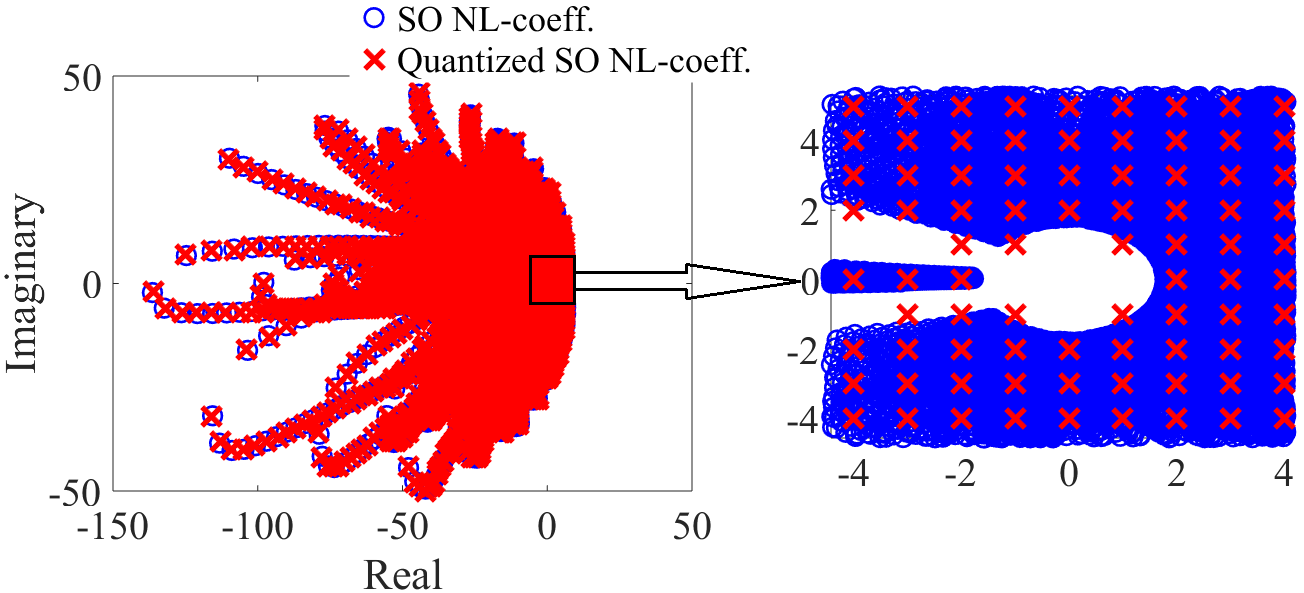}
		\par\end{centering}
	\centering{}\caption{{\footnotesize{}Quantized SO nonlinearity coefficients for a transmission distance of 2800 km.}}
\end{figure}

Fig. 9 shows the results of the coefficient quantization for the SO perturbative nonlinear coefficients. It is clear from Fig. 9 that the number of distinct nonlinearity coefficients is dramatically reduced after the quantization process since several of the similar coefficients are approximated to the same quantized coefficient.  Fig. 10 illustrates the number of SO coefficients kept after truncation, quantization, and combination when compared to the actual number of coefficients as a function of the transmission distance for a fixed window size of $L_{w}=100$.
\begin{figure}[H]
	\centering{}\includegraphics[width=1\columnwidth,height=0.17\paperheight]{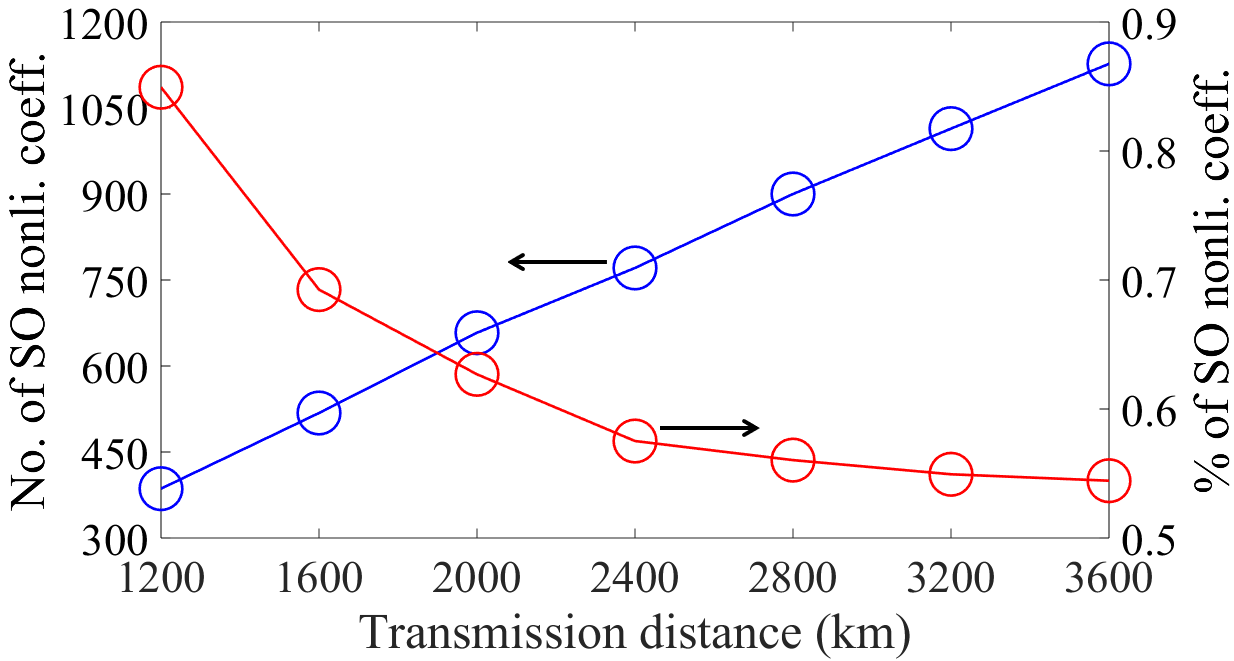}\caption{{Number of SO nonlinearity coefficients kept after truncation, quantization, and combination (blue curve) as a function of the transmission distance. The red curve shows the percentage of coefficients retained after truncation, quantization, and combination.}}
\end{figure}

The results in Fig. 10 indicate that the number of SO coefficients kept after truncation, quantization, and combination linearly increases as the transmission distance increases. However, the percentage of the SO coefficients over the total number of coefficients exponentially decreases as the transmission distance increases for a fixed window size of $L_{w}=100.$ For example, for a transmission distance of $2800$ km, the number of SO perturbation terms with a coefficient truncation threshold of $\mu=-40$ dB is $160606$, which is reduced to $900$ after the coefficient quantization and combining the perturbation terms that have the same coefficients, as in \cite{ZTao2013}. In other words, only $0.56 \%$ of coefficients are kept after the truncation, quantization, and combination to carry out the SO distortion field calculation, which significantly reduces the implementation complexity of the SO-PB-NLC technique. 

\begin{figure*}[!t]
	\noindent\begin{minipage}[t]{1\columnwidth}%
		\includegraphics[width=1\columnwidth,height=0.17\paperheight]{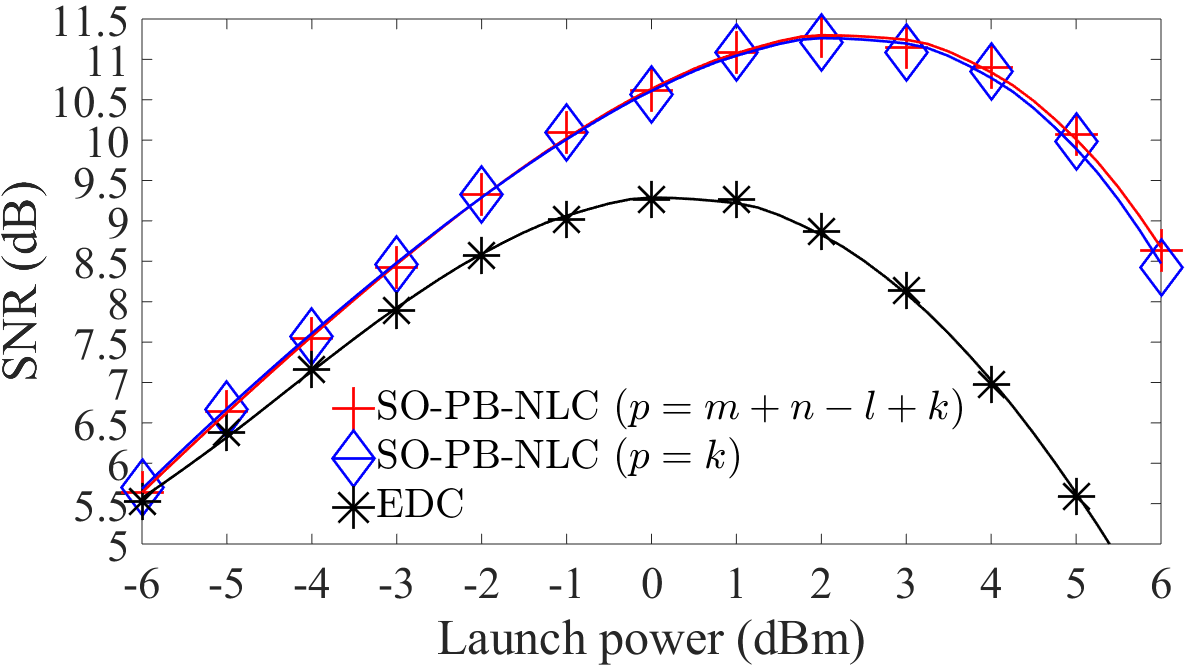}
		\begin{center}
			\vspace{-0.3cm}
			\,\,\,\,\,\,\,\,\,\,\,(a)
			\par\end{center}%
	\end{minipage}%
	\vspace{0.1cm}
	\noindent\begin{minipage}[t]{1\columnwidth}%
		\includegraphics[width=1\columnwidth,height=0.17\paperheight]{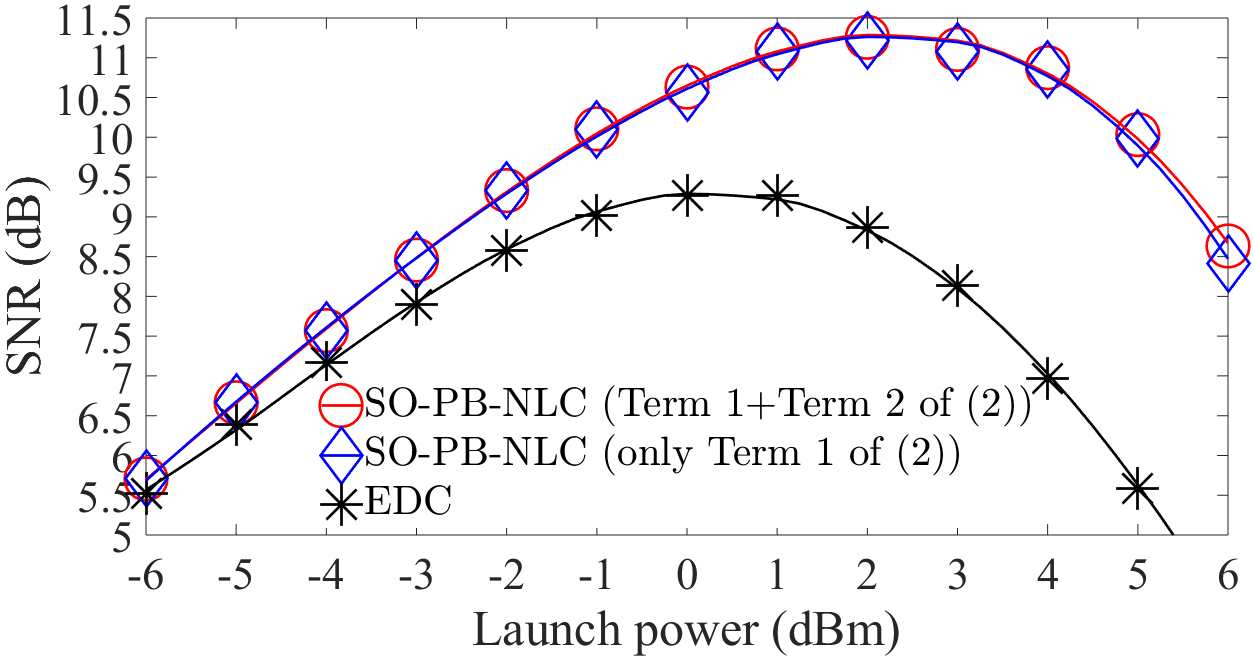}%
		\begin{center}
			\vspace{-0.3cm}
			\,\,\,\,\,\,\,\,\,\,\,(b)
			\par\end{center}%
	\end{minipage}\\
	
	\centering{}\caption{{(a) SNR vs. launch power with $p=m+n-l+k$ and $p=k$; (b) SNR vs. launch power with Term 1+Term 2 and only Term 1 of (1).}}
	\vspace{0.7cm}
\end{figure*} 

\begin{figure*}[!t]
	
	\noindent\begin{minipage}[t]{1\columnwidth}%
		\includegraphics[width=1\columnwidth,height=0.17\paperheight]{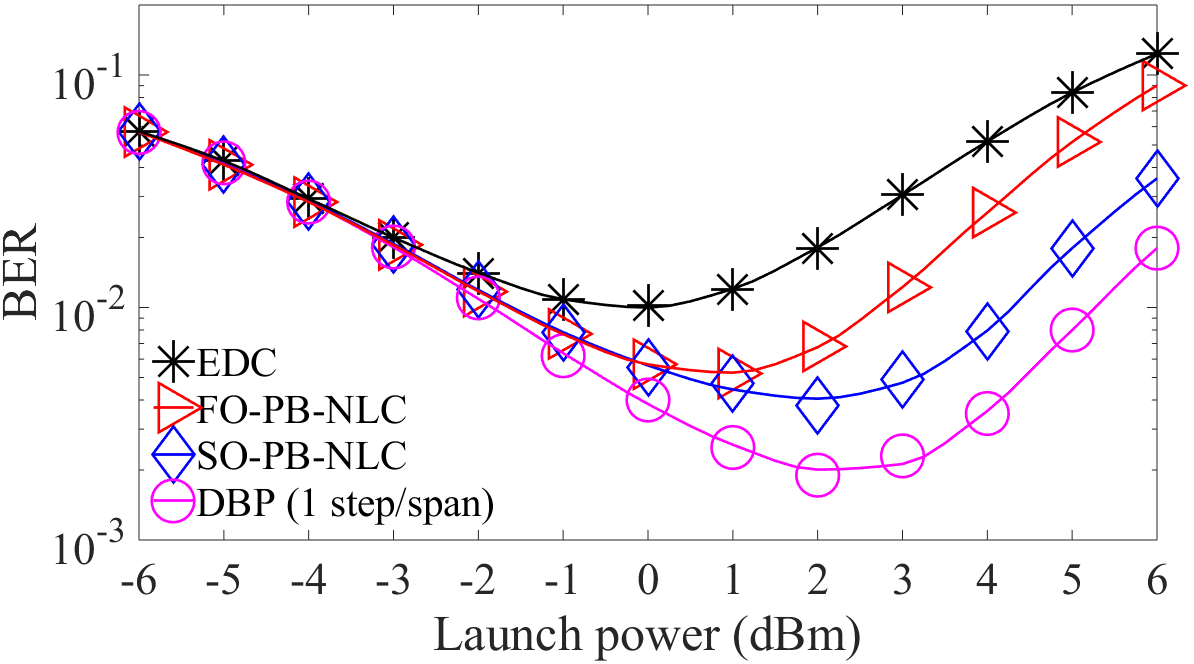}%
		\begin{center}
			\vspace{-0.3cm}
			\,\,\,\,\,\,\,\,\,\,\,(a)
			\par\end{center}%
	\end{minipage}%
	\noindent\begin{minipage}[t]{1\columnwidth}%
		\includegraphics[width=1\columnwidth,height=0.17\paperheight]{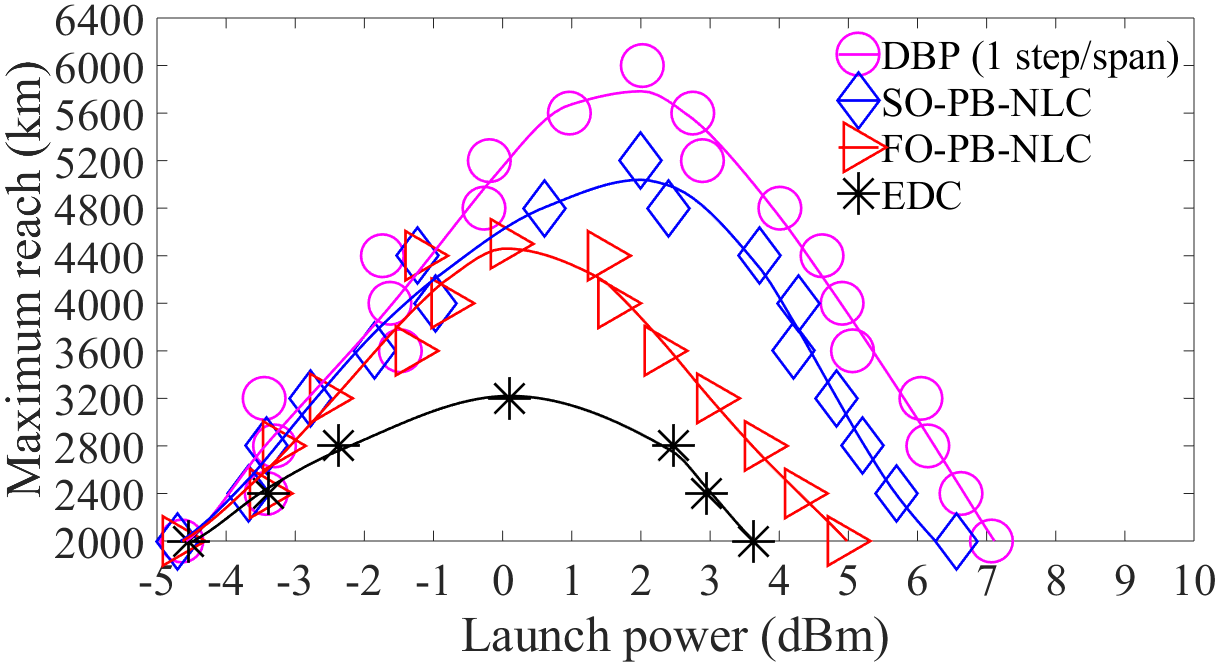}%
		\begin{center}
			\vspace{-0.3cm}
			\,\,\,\,\,\,\,\,\,\,\,(b)
			\par\end{center}%
	\end{minipage}
	\centering{}\caption{{(a) BER vs. launch power; and (b) Maximum reach vs. launch power.}}
	\vspace{0.7cm}
\end{figure*} 

\section{Numerical Simulations and Discussions}

We consider a single-channel Pol-Mux optical transmission system and the SO-PB-NLC technique is applied as a predistortion at the transmitter. After RRC pulse shaping, the predistorted signal is up-converted to optical domain and transmitted over the long-haul optical fiber link.  The simulation parameters used for the study are listed in Table 1. 
\begin{table}[H]
	\renewcommand{\arraystretch}{1.2}
	\caption{Simulation Parameters \cite{Liang2014}, \cite{OSSunish Kumar2019}, \cite{Tao2011}, \cite{S. T. Le2015PCSC}.} 
	\centering
	\begin{tabular}{|c|c|}
		\hline
		Parameter & Value\\
		\hline
		\hline
		RRC filter roll-off factor & $0.1$ \\
		\hline
		$\mu$ & $-40$ dB \\
		\hline
		Fiber span length & $80$ km\\
		\hline
		$\alpha$ & $0.2$ dB/km\\
		\hline 
		$\beta_{2}$ & $-20.47$  $\textrm{p\ensuremath{\textrm{s}^{2}}}$/km\\
		\hline 
		$\gamma$ & $1.22$  (1/W)/km\\
		\hline 
		$\textrm{Polarization mode dispersion coefficient}$ & $0.1$  $\textrm{ps/\ensuremath{\sqrt{\textrm{km}}}}$\\
		\hline 
		$\textrm{Noise figure of EDFA }$ & $5.5$  dB\\
		\hline  
	\end{tabular}
\end{table}
The modulation format considered is 16-quadrature amplitude modulation. The data transmission rate is 32 Gbaud. The polarization state, carrier phase, and symbol timing are assumed perfectly known at the receiver \cite{NVIrukulapati2014}. The amplified spontaneous emission (ASE) noise of EDFA is added to the signal after each fiber span to capture the nonlinear interaction between the signal and the ASE noise \cite{OSSunish Kumar2019}.

\subsection{Simulation Results}

It is observed from Fig. 11(a) that neglecting the FO ghost pulse at arbitrary time indices $m+n-l\neq0$ only slightly affects the received SNR of the SO-PB-NLC technique. That is because the CD-induced pulse overlap between the FO ghost pulse at arbitrary time indices $m+n-l\neq0$ and the zeroth-order pulses is significantly less, and thereby, the magnitude of the corresponding nonlinearity coefficient is negligibly small. Similarly, results in Fig. 11(b) show that the received SNR of the SO-PB-NLC is only slightly affected when Term 2 of (\ref{eqn2}) is neglected. Based on these considerations, we select the implementation of the SO-PB-NLC technique with phase-matching conditions $l=m+n$ and $p=k$ and by taking into account only Term 1 of (\ref{eqn2}) for further numerical investigations. 

Fig. 12(a) shows BER as a function of the launch power for the single-channel Pol-Mux optical transmission system at a transmission distance of 2800 km. In Fig. 12(a), the performance of the DBP technique is included along with the FO-PB-NLC and the electronic dispersion compensation (EDC) techniques for comparison. It is worth mentioning that, in this work, we chose the conventional DBP technique as a benchmark to compare the performance with our proposed SO-PB-NLC technique. The transmission distance considered is 2800 km. We observe from Fig. 12(a) that the BER performance of the SO-PB-NLC technique is significantly better than that of the FO-PB-NLC and EDC techniques. It is observed that the optimal launch power for the SO-PB-NLC technique is increased by $\sim2$ dB and $\sim1$ dB when compared to the EDC and FO-PB-NLC techniques, respectively. Another observation is that the BER performance of the DBP technique is higher than that of the proposed SO-PB-NLC technique. That is because the DBP is a numerical method that uses the SSFM, and so it compensates for the nonlinearity effects span-by-span \cite{Ip2008}. On the other hand, the PB-NLC techniques use an analytical approximation for the solution of the NLSE with the assumption that the fiber link has only one span \cite{Tao2011}. That is a general assumption considered in the design of the PB-NLC techniques. It is important to mention that the single span assumption of the PB-NLC techniques allows the compensation of the nonlinearity effect in a single computation step, thus reducing the computational effort required \cite{Tao2011}. 

Fig. 12(b) shows the plot of the maximum transmission reach as a function of the launch power for DBP, SO-PB-NLC, FO-PB-NLC, and EDC techniques at 20\% overhead soft-decision forward error-correction limit with a BER value of $2\times10^{-2}$ \cite{LBomin2014} for a single-channel Pol-Mux optical transmission system. It is observed that the maximum transmission reach for DBP, SO-PB-NLC, FO-PB-NLC, and EDC is 6000 km, 5200 km, 4500 km, and 3200 km, respectively. These results indicate that the SO-PB-NLC technique provides an extended transmission reach by 62.5\% and 15.6\% when compared to EDC and the FO-PB-NLC techniques, respectively. Further, it is also observed that the nonlinearity threshold of the SO-PB-NLC technique is improved by $\sim4.7$ dB and $\sim1.7$ dB when compared to the EDC and FO-PB-NLC techniques, respectively. 
\subsection{Complexity Evaluation} 

In this section, the computational complexities of the DBP, SO-PB-NLC, FO-PB-NLC, and EDC techniques are evaluated based on the number of real-valued multiplications per symbol for the Pol-Mux optical transmission system. It is important to mention that the nonlinearity coefficient matrices of the FO-/SO-PB-NLC techniques are truncated at a threshold of $\mu$=-40 dB \cite{Tao2011}. Also, in the implementation, the nonlinearity coefficient matrices are quantized according to the method given in \cite{ZTao2013}. For DBP with $N_{steps}$ per span, the expression for the number of real-valued multiplications per symbol is given as $8N_{steps}N_{spans}N_{\textrm{FFT}}(\log_{2}(N_{\textrm{FFT}})+10.5)/N_{s},$ where $N_{spans}$ is the number of fiber spans, $N_{\textrm{FFT}}$ is the fast Fourier transform size, and $N_{s}$ is the number of samples \cite{Amari2017}. In case of the FO-/SO-PB-NLC techniques, the triplet/quintuplet symbols in the nonlinear distortion calculation can be stored in LUT; therefore, the number of real-valued multiplications per symbol can be represented as $2(4M+3)$ \cite{Liang2014}, where $M$ is the number of significant perturbation coefficients in the nonlinearity coefficient matrix $\left.\textbf{C}_{m,n}^{\text{FO}}\right/$$\widetilde{\textbf{C}}_{m,n,k}^{\text{SO,\,Term\,1}}$. It is worth noting that the value of $M$ increases with increasing the number of fiber spans because of the corresponding increase in the number of coefficients in the nonlinearity coefficient matrix, satisfying the truncation threshold. For the EDC technique, the number of real-valued multiplications per symbol is given as $8N_{\textrm{FFT}}(\log_{2}(N_{\textrm{FFT}})+1)/N_{s}$ \cite{Amari2017}.  

\begin{figure}[h]
	\centering{}\includegraphics[width=1\columnwidth,height=0.17\paperheight]{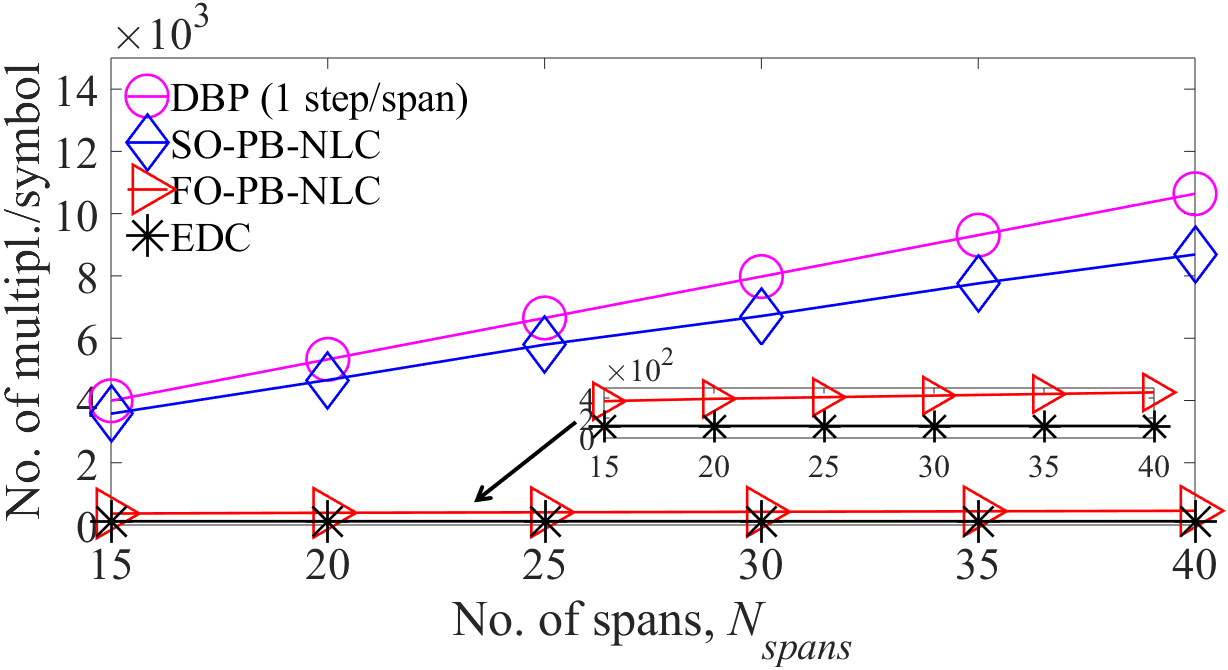}\caption{{Number of real-valued multiplications/symbol for DBP, SO-PB-NLC, FO-PB-NLC, and EDC techniques as a function of the number of fiber spans $N_{spans}.$}}
\end{figure}

Fig. 13 depicts the number of real-valued multiplications per symbol as a function of the number of fiber spans, $N_{spans}$ for DBP, SO-PB-NLC, FO-PB-NLC, and EDC techniques. The results indicate that the complexity of the DBP technique increases rapidly as $N_{spans}$ increases, which is attributed to the corresponding increase in the computation steps for the SSFM technique \cite{Amari2017}. On the other hand, the complexity increase for the FO-/SO-PB-NLC techniques is due to the increase in the number of quantized nonlinearity coefficients as the number of fiber span increases. The result in Fig. 13 shows that the computational complexity of the proposed SO-PB-NLC technique is less than that of the DBP technique. For example, at $N_{spans}$= 35 (i.e., at 2800 km), the required number of real-valued multiplications for the SO-PB-NLC technique is $1550$ fewer than that of the DBP technique with 1 step/span. It is important to note that the nonlinearity coefficients of the FO-/SO-PB-NLC techniques can be calculated beforehand and stored in LUT. That is the key factor that makes the proposed SO-PB-NLC technique less complex than the conventional DBP technique. 

\begin{figure}[H]
	\centering{}\includegraphics[width=1\columnwidth,height=0.17\paperheight]{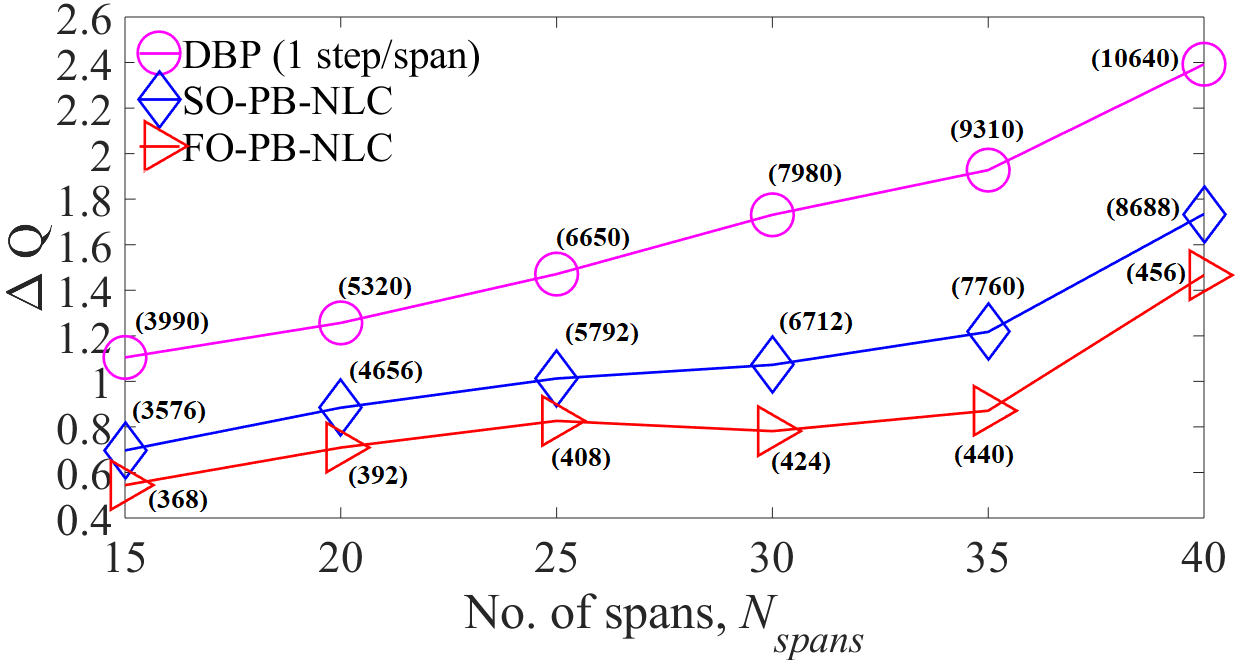}\caption{{$Q$-factor gain as a function of the number of fiber spans and the computational complexity.}}
\end{figure}

Fig. 14 shows the plot of the $Q$-factor gain $\Delta Q$, defined as $\Delta Q = Q_{FO-PB-NLC/SO-PB-NLC/DBP}-Q_{EDC}$, as a function of the number of fiber spans. The corresponding computational complexity value is shown in parentheses. Fig. 14 indicates that even though the proposed SO-PB-NLC technique is computationally efficient when compared with DBP, its implementation complexity is large when compared to FO-PB-NLC. While the SO-PB-NLC approach appears to be complex, we believe it would be possible to further reduce the complexity by using dimensionality reduction techniques, such as the principal component analysis. This represents the focus of future work.  

\section{Conclusion}
In this paper, we have designed the SO-PB-NLC technique for single-channel Pol-Mux systems. We have shown through numerical simulations that the NLC performance of the designed SO-PB-NLC technique is improved in comparison with the FO-PB-NLC technique. We have demonstrated that the SO-PB-NLC technique provides an extended transmission reach compared with the EDC and FO-PB-NLC, respectively, for a single-channel Pol-Mux optical link. We have also shown that the computational burden of the SO-PB-NLC technique is lower than that of the DBP technique with one step per span. 

\appendices
\section{Proof of Lemma 1}

The differential equation governing the SO distortion field for the Pol-Mux signal by considering only Term 1 of the nonlinear part in (\ref{eqn2}) can be represented as:

\begin{multline}
	\label{eqn24}
	\frac{\partial}{\partial z}u_{2,x/y}(z,t)=-j\frac{\beta_{2}}{2}\frac{\partial^{2}}{\partial t^{2}}u_{2,x/y}(z,t)+j2\tilde{u}_{1,x/y}(z,t)\\
	\left(\left|u_{0,x/y}(z,t)\right|^{2}+\left|u_{0,y/x}(z,t)\right|^{2}\right)\exp(-\alpha z).
\end{multline}

By taking the Fourier transform of (\ref{eqn24}) and integrating with respect to $z$ from $0$ to $L$ with the assumption of an ideal dispersion compensation at $z=L$, we get the solution in frequency-domain as:

\begin{multline}
	\label{eqn25}
	U_{2,x/y}^{\text{Term\,1}}(L,w)=\frac{64}{81}\gamma^{2}\intop_{0}^{L}2F_{x/y}^{\text{Term\,1}}(z,w)\\
	\times\exp\left(-j\frac{w^{2}\beta_{2}z}{2}\right)
	\exp(-\alpha z)dz.
\end{multline} 
Note that $\frac{64}{81}\gamma^{2}$ in (\ref{eqn25}) is the nonlinearity scaling factor for the SO distortion field that is obtained from the perturbation series representation of the output optical field, as shown in \cite{Vannucci2002}. The $F_{x/y}^{\text{Term\,1}}(z,w)$ in (\ref{eqn25}) is given as:
\begin{multline}
	\label{eqn26}
	F_{x/y}^{\text{Term\,1}}(z,w)=\intop_{-\infty}^{\infty}\left(\left|u_{0,x/y}(z,t)\right|^{2}+\left|u_{0,y/x}(z,t)\right|^{2}\right)\\
	\times\tilde{u}_{1,x/y}(z,t)\exp\left(-jwt\right)dt.
\end{multline}
By substituting (\ref{eqn3}) in (\ref{eqn26}), the equation for $F_{x/y}^{\text{Term\,1}}(z,w)$ can be represented as: 
\vspace{-0.3cm}
\begin{multline}
	\label{eqn27}
	F_{x/y}^{\text{Term\,1}}(z,w)=P_{0}^{5/2}\sum_{m}\sum_{n}\sum_{l}\sum_{k}\sum_{p}\\
	\left(a_{m,x/y}a_{l,x/y}^{*}+a_{m,y/x}a_{l,y/x}^{*}\right)a_{n,x/y}\\
	\left(a_{k,x/y}a_{p,x/y}^{*}+a_{k,y/x}a_{p,y/x}^{*}\right)
	\intop_{-\infty}^{\infty}\tilde{g}_{1,m+n-l}(z,s,t-\\
	(m+n-l)T)\hat{g}_{k}(z,t-kT)\hat{g}_{p}^{*}(z,t-pT) \exp\left(-jwt\right) dt.
\end{multline}
Next, substituting the simplifying assumptions $l=m+n$ and $p=k$ in (\ref{eqn27}), we get: 
\vspace{-0.3cm}
\begin{multline}
	\label{eqn28}
	F_{x/y}^{\text{Term\,1}}(z,w)=P_{0}^{5/2}\sum_{m}\sum_{n}\sum_{k}\left(a_{m,x/y}a_{m+n,x/y}^{*}\right.\\
	\left.+a_{m,y/x}a_{m+n,y/x}^{*}\right)a_{n,x/y}\left(a_{k,x/y}a_{k,x/y}^{*}+a_{k,y/x}a_{k,y/x}^{*}\right)\\
	\times\intop_{-\infty}^{\infty}\tilde{g}_{1,0}(z,s,t)\hat{g}_{k}(z,t-kT)\hat{g}_{k}^{*}(z,t-kT)\exp\left(-jwt\right)dt.
\end{multline}
In (\ref{eqn28}), $\tilde{g}_{1,0}(z,s,t)$ is the FO distortion field at the zeroth time index affected by CD, which is obtained by the convolution between the temporal CD term $\frac{1}{\sqrt{-2\pi j\beta_{2}z}}\exp\left(\frac{-jt^{2}}{2\beta_{2}z}\right)$ and the FO nonlinearity coefficient in \cite{Tao2011}. In our analysis, we consider Gaussian shape assumption for the input pulses, i.e., $\hat{g}(z=0,t)=\exp\left(-\frac{t^{2}}{2\tau^{2}}\right),$ where $\tau$ represents the pulse width. With this Gaussian shape assumption, the FO distortion field $\tilde{g}_{1,0}(z,s,t)$ can be represented as:  
\vspace{-0.2cm}
\begin{multline}
	\label{eqn29}
	\tilde{g}_{1,0}(z,s,t)=\frac{1}{\sqrt{-2\pi j\beta_{2}z}}\exp\left(\frac{-jt^{2}}{2\beta_{2}z}\right)\\
	\otimes\left(\vphantom{\exp\left\{ \begin{array}{c}
			-\frac{3\left[2t/3+T_{m}-T_{l}\right]\left[2t/3+T_{n}-T_{l}\right]}{\tau^{2}(1+3j\beta_{2}s/\tau^{2})}\\
			-\frac{(T_{n}-T_{m})^{2}}{\tau^{2}\left[1+2j\beta_{2}s/\tau^{2}+3(\beta_{2}s/\tau^{2})^{2}\right]}
		\end{array}\right\} }\exp(\frac{-t^{2}}{6\tau^{2}})\intop_{0}^{z}\frac{\exp(-\alpha s)}{\sqrt{1+2j\beta_{2}s/\tau^{2}+3(\beta_{2}s/\tau^{2})^{2}}}\right.\\
	\times\left.\exp\left\{ \begin{array}{c}
		-\frac{3\left[2t/3-nT\right]\left[2t/3-mT\right]}{\tau^{2}(1+3j\beta_{2}s/\tau^{2})}\\
		-\frac{(n-m)^{2}T^{2}}{\tau^{2}\left[1+2j\beta_{2}s/\tau^{2}+3(\beta_{2}s/\tau^{2})^{2}\right]}
	\end{array}\right\} ds\vphantom{\exp(\frac{-t^{2}}{6\tau^{2}})\intop_{0}^{z}\frac{\exp(-\alpha s)}{\sqrt{1+2j\beta_{2}s/\tau^{2}+3(\beta_{2}s/\tau^{2})^{2}}}}\right)\\
	=j\tau^{3}\intop_{0}^{z}\frac{\exp(-\alpha s)}{\sqrt{-j\overline{B}(s)\widehat{D}(z,s)}}\exp\left\{ \frac{-j}{\overline{B}(s)\widehat{D}(z,s)}\right.\\
	\left[(\ddddot{A}_{m,n}T^{2}-2t\ddddot{B}_{m,n}T+\frac{3}{2}t^{2})\tau^{4}-j\beta_{2}((m^{2}z\right.\\
	+(3s-z)mn+n^{2}z)T^{2}+2st\ddddot{C}_{m,n}T+st^{2})\tau^{2}\\
	\left.\left.-(mnzT^{2}-\frac{1}{2}st^{2})\beta_{2}^{2}s\right]\vphantom{\frac{-j}{\overline{B}(s)\widehat{D}(z,s)}}\right\} ds,
\end{multline}
where $\ddddot{A}_{m,n}=(m^{2}+mn+n^{2}),\, {\displaystyle \ddddot{B}_{m,n}=(m+n)},$ and ${\displaystyle \ddddot{C}_{m,n}=({\displaystyle m+n-2}).}$\\
Next, substituting (\ref{eqn29}) in (\ref{eqn28}) and the resultant equation in (\ref{eqn25}), we obtain the SO nonlinear distortion field corresponding to Term 1 with the simplifying assumptions $l=m+n$ and $p=k$ as:
\begin{multline}
	\label{eqn30}
	\widetilde{U}_{2,x/y}^{\text{Term\,1}}(L,w)=\frac{64}{81}\gamma^{2}P_{0}^{5/2}\sum_{m}\sum_{n}\sum_{k}\\
	2\left(a_{m,x/y}a_{m+n,x/y}^{*}+a_{m,y/x}a_{m+n,y/x}^{*}\right)a_{n,x/y}\\
	\times\left(a_{k,x/y}a_{k,x/y}^{*}+a_{k,y/x}a_{k,y/x}^{*}\right)\widetilde{G}_{m,n,k}^{\text{Term\,1}}(z,s,w),
\end{multline} 
where $\widetilde{G}_{m,n,k}^{\text{Term\,1}}(z,s,w)$ is given by:
\begin{multline}
	\label{eqn31}
	\widetilde{G}_{m,n,k}^{\text{Term\,1}}(z,s,w)=\intop_{0}^{L}\exp(-\alpha z)\left(\intop_{-\infty}^{\infty}\tilde{g}_{1,0}(z,s,t)\right.\\
	\left.\hat{g}_{k}(z,t-kT)\hat{g}_{k}^{*}(z,t-kT)\exp\left(-jwt\right)\,dt\vphantom{\intop_{-\infty}^{\infty}\tilde{g}_{1,0}(z,t)}\right)\\
	\times\exp\left(-j\frac{w^{2}\beta_{2}z}{2}\right)dz.
\end{multline}  
It is clear from (\ref{eqn30}) that the function $\widetilde{G}_{m,n,k}^{\text{Term\,1}}(z,s,w)$ calculates the coefficient of nonlinear interaction between the quintuplet pulses in frequency-domain. The corresponding time-domain nonlinearity coefficient is obtained by taking the inverse Fourier transform of (\ref{eqn31}), which can be represented as:
\vspace{-0.2cm}
\begin{multline}
	\label{eqn32}
	\widetilde{g}_{m,n,k}^{\text{Term\,1}}(z,s,t)=\intop_{0}^{L}\exp(-\alpha z)\\
	\left(\vphantom{{\displaystyle \frac{1}{\sqrt{-2\pi j\beta_{2}z}}\exp\left(\frac{-jt^{2}}{2\beta_{2}z}\right)}}\tilde{g}_{1,0}(z,s,t)\right.
	\left.\hat{g}_{k}(z,t-kT)\hat{g}_{k}^{*}(z,t-kT)\vphantom{{\displaystyle \frac{1}{\sqrt{-2\pi j\beta_{2}z}}\exp\left(\frac{-jt^{2}}{2\beta_{2}z}\right)}}\right)\\
	\otimes\left({\displaystyle \frac{1}{\sqrt{-2\pi j\beta_{2}z}}\exp\left(\frac{-jt^{2}}{2\beta_{2}z}\right)}\right)dz.
\end{multline} 
Next, substituting the expression for zeroth-order (or linearly dispersed) pulse $\hat{g}_{k}(z,t-kT)=\frac{\tau}{\sqrt{\tau^{2}-j\beta_{2}z}}\exp\left(\frac{\left(kT-t\right)^{2}}{2(j\beta_{2}z-\tau^{2})}\right)$ and (\ref{eqn29}) in (\ref{eqn32}) with the assumption of a symbol rate operation (i.e., $t=0$), we obtain (\ref{eqn6}), i.e.,
\begin{equation}
	\label{eqn33}
	\widetilde{\textbf{C}}_{m,n,k}^{\text{SO,\,Term\,1}}=\left.\widetilde{g}_{m,n,k}^{\text{Term\,1}}(z,s,t)\right|_{t=0}.
\end{equation}
\section{Proof of Lemma 2}

By considering Term 2 of the nonlinear part in (\ref{eqn2}), the propagation equation governing the evolution of the SO distortion field can be represented as:

\begin{multline}
	\label{eqn34}
	\frac{\partial}{\partial z}u_{2,x/y}(z,t)=-j\frac{\beta_{2}}{2}\frac{\partial^{2}}{\partial t^{2}}u_{2,x/y}(z,t)+j\exp(-\alpha z)\\
	\times\left(u_{0,x/y}^{2}(z,t)+u_{0,y/x}^{2}(z,t)\right)\tilde{u}_{1,x/y}^{*}(z,t).
\end{multline}
The solution of (\ref{eqn34}) in frequency-domain can be obtained as:

\begin{multline}
	\label{eqn35}
	U_{2,x/y}^{\text{Term\,2}}(L,w)=\frac{64}{81}\gamma^{2}\intop_{0}^{L}F_{x/y}^{\text{Term\,2}}(z,w)
	\exp\left(-j\frac{w^{2}\beta_{2}z}{2}\right)\\
	\times\exp(-\alpha z)dz,
\end{multline} 
where $F_{x/y}^{\text{Term\,2}}(z,w)$ is given as:

\begin{multline}
	\label{eqn36}
	F_{x/y}^{\text{Term\,2}}(z,w)=\intop_{-\infty}^{\infty}\left(u_{0,x/y}^{2}(z,t)+u_{0,y/x}^{2}(z,t)\right)\\
	\times\tilde{u}_{1,x/y}^{*}(z,t)\exp\left(-jwt\right)dt.
\end{multline}
By substituting (\ref{eqn3}) in (\ref{eqn36}), we obtain:

\begin{multline}
	\label{eqn37}
	F_{x/y}^{\text{Term\,2}}(z,w)=P_{0}^{5/2}\sum_{m}\sum_{n}\sum_{l}\sum_{k}\sum_{p}a_{n,x/y}^{*}\\
	\left(a_{m,x/y}^{*}a_{l,x/y}+a_{m,y/x}^{*}a_{l,y/x}\right)\left(\vphantom{a_{m,x/y}^{*}}a_{k,x/y}a_{p,x/y}+a_{k,y/x}a_{p,y/x}\right)\\
	\times\intop_{-\infty}^{\infty}\tilde{g}_{1,m+n-l}^{*}(z,s,t-(m+n-l)T)\\
	\times\hat{g}_{k}(z,t-kT)\hat{g}_{p}(z,t-pT)\exp\left(-jwt\right)dt.
\end{multline}
Next, substituting the simplifying assumptions $l=m+n$ and $p=-k$ in (\ref{eqn37}), we get:

\begin{multline}
	\label{eqn38}
	F_{x/y}^{\text{Term\,2}}(z,w)=P_{0}^{5/2}\sum_{m}\sum_{n}\sum_{k}\\
	\left(a_{m,x/y}^{*}a_{m+n,x/y}+a_{m,y/x}^{*}a_{m+n,y/x}\right)a_{n,x/y}^{*}\\
	\left(\vphantom{a_{m,x/y}^{*}}a_{k,x/y}a_{-k,x/y}+a_{k,y/x}a_{-k,y/x}\right)
	\intop_{-\infty}^{\infty}\tilde{g}_{1,0}^{*}(z,s,t)\hat{g}_{k}(z,t-kT)\\
	\times\hat{g}_{k}(z,t-kT)\exp\left(-jwt\right)dt.
\end{multline}

Next, substituting (\ref{eqn29}) in (\ref{eqn38}) and the resultant equation in (\ref{eqn35}), we obtain the SO nonlinear distortion field corresponding to Term 2 with the simplifying assumptions $l=m+n$ and $p=-k$ as:

\begin{multline}
	\label{eqn39}
	\widetilde{U}_{2,x/y}^{\text{Term\,2}}(L,w)=\frac{64}{81}\gamma^{2}P_{0}^{5/2}\sum_{m}\sum_{n}\sum_{k}\\
	\left(a_{m,x/y}^{*}a_{m+n,x/y}+a_{m,y/x}^{*}a_{m+n,y/x}\right)a_{n,x/y}^{*}\\
	\times\left(a_{k,x/y}a_{-k,x/y}+a_{k,y/x}a_{-k,y/x}\right)\widetilde{G}_{m,n,k}^{\text{Term\,2}}(z,s,w),
\end{multline} 
where $\widetilde{G}_{m,n,k}^{\text{Term\,2}}(z,s,w)$ is given by:
\begin{multline}
	\label{eqn40}
	\widetilde{G}_{m,n,k}^{\text{Term\,2}}(z,s,w)=\intop_{0}^{L}\exp(-\alpha z)\left(\intop_{-\infty}^{\infty}\tilde{g}_{1,0}^{*}(z,s,t)\right.\\
	\left.\hat{g}_{k}(z,t-kT)\hat{g}_{k}(z,t-kT)\exp\left(-jwt\right)\,dt\vphantom{\intop_{-\infty}^{\infty}\tilde{g}_{1,0}(z,t)}\right)\\
	\times\exp\left(-j\frac{w^{2}\beta_{2}z}{2}\right)dz.
\end{multline}
The time-domain nonlinearity coefficient is obtained by taking the inverse Fourier transform of (\ref{eqn40}), which can be represented as:

\begin{multline}
	\label{eqn41}
	\widetilde{g}_{m,n,k}^{\text{Term\,2}}(z,s,t)=\intop_{0}^{L}\exp(-\alpha z)\\
	\left(\vphantom{{\displaystyle \frac{1}{\sqrt{-2\pi j\beta_{2}z}}\exp\left(\frac{-jt^{2}}{2\beta_{2}z}\right)}}\tilde{g}_{1,0}^{*}(z,s,t)\right.
	\left.\hat{g}_{k}(z,t-kT)\hat{g}_{k}(z,t-kT)\vphantom{{\displaystyle \frac{1}{\sqrt{-2\pi j\beta_{2}z}}\exp\left(\frac{-jt^{2}}{2\beta_{2}z}\right)}}\right)\\
	\otimes\left({\displaystyle \frac{1}{\sqrt{-2\pi j\beta_{2}z}}\exp\left(\frac{-jt^{2}}{2\beta_{2}z}\right)}\right)dz.
\end{multline} 
Substituting $\hat{g}_{k}(z,t-kT)=\frac{\tau}{\sqrt{\tau^{2}-j\beta_{2}z}}\exp\left(\frac{\left(kT-t\right)^{2}}{2(j\beta_{2}z-\tau^{2})}\right)$ and (\ref{eqn29}) in (\ref{eqn41}) with the assumption of a symbol rate operation (i.e., $t=0$), we obtain (\ref{eqn14}), i.e.,

\begin{equation}
	\label{eqn42}
	\widetilde{\textbf{C}}_{m,n,k}^{\text{SO,\,Term\,2}}=\left.\widetilde{g}_{m,n,k}^{\text{Term\,2}}(z,s,t)\right|_{t=0}.
\end{equation}

\section{Proof of Theorem 1}

The solution of (\ref{eqn2}) in frequency-domain with the simplifying assumptions $l=m+n$ and $p=k$ for Term 1 and $l=m+n$ and $p=-k$ for Term 2 can be represented as:   

\begin{multline}
	\label{eqn43}
	\widetilde{U}_{2,x/y}(L,w)=\frac{64}{81}\gamma^{2}\intop_{0}^{L}F_{x/y}(z,w)\\
	\times\exp\left(-j\frac{w^{2}\beta_{2}z}{2}\right)
	\exp(-\alpha z)dz,
\end{multline} 
where 
\begin{equation}
	\label{eqn44}
	F_{x/y}(z,w)=2F_{x/y}^{\text{Term\,1}}(z,w)+F_{x/y}^{\text{Term\,2}}(z,w).
\end{equation}
By substituting (\ref{eqn28}) and (\ref{eqn38}) in (\ref{eqn44}), we get:

\begin{multline}
	\label{eqn45}
	F_{x/y}(z,w)=P_{0}^{5/2}\sum_{m}\sum_{n}\sum_{k}\\
	\left[\vphantom{\intop_{-\infty}^{\infty}\tilde{g}_{1,0}^{*}(z,s,t)}2\left(a_{m,x/y}a_{m+n,x/y}^{*}+a_{m,y/x}a_{m+n,y/x}^{*}\right)a_{n,x/y}\right.\\
	\times\left(a_{k,x/y}a_{k,x/y}^{*}+a_{k,y/x}a_{k,y/x}^{*}\right)\\
	\intop_{-\infty}^{\infty}\tilde{g}_{1,0}(z,s,t)\hat{g}_{k}(z,t-kT)
	\times\hat{g}_{k}^{*}(z,t-kT)\exp\left(-jwt\right)dt\\
	+\left(a_{m,x/y}^{*}a_{m+n,x/y}+a_{m,y/x}^{*}a_{m+n,y/x}\right)a_{n,x/y}^{*}\\
	\times \left(a_{k,x/y}a_{-k,x/y}+a_{k,y/x}a_{-k,y/x}\right)\intop_{-\infty}^{\infty}\tilde{g}_{1,0}^{*}(z,s,t)\\
	\left.\times\hat{g}_{k}(z,t-kT)\hat{g}_{k}(z,t-kT)\exp\left(-jwt\right)dt\vphantom{\intop_{-\infty}^{\infty}\tilde{g}_{1,0}^{*}(z,s,t)}\right].
\end{multline}

Next, substituting (\ref{eqn45}) in (\ref{eqn43}) and substituting the expressions for the FO ghost pulse and the linearly dispersed pulses, we obtain the SO distortion field as:

\begin{multline}
	\label{eqn46}
	\widetilde{U}_{2,x/y}(L,w)=\frac{64}{81}\gamma^{2}P_{0}^{5/2}\sum_{m}\sum_{n}\sum_{k}\\
	\left[\vphantom{\sum}2\left(a_{m,x/y}a_{m+n,x/y}^{*}+a_{m,y/x}a_{m+n,y/x}^{*}\right)a_{n,x/y}\right.\\
	\times\left(a_{k,x/y}a_{k,x/y}^{*}+a_{k,y/x}a_{k,y/x}^{*}\right)\widetilde{G}_{m,n,k}^{\text{Term\,1}}(z,s,w)\\
	+\left(a_{m,x/y}^{*}a_{m+n,x/y}+a_{m,y/x}^{*}a_{m+n,y/x}\right)a_{n,x/y}^{*}\\
	\left.\times \left(a_{k,x/y}a_{-k,x/y}+a_{k,y/x}a_{-k,y/x}\right)\widetilde{G}_{m,n,k}^{\text{Term\,2}}(z,s,w)\vphantom{\sum}\right],
\end{multline}
where $\widetilde{G}_{m,n,k}^{\text{Term\,1}}(z,s,w)$ and $\widetilde{G}_{m,n,k}^{\text{Term\,2}}(z,s,w)$ are given by (\ref{eqn31}) and (\ref{eqn40}), respectively.

By calculating the inverse Fourier transform of (\ref{eqn46}) and assuming the symbol rate operation (i.e., $t=0$), we obtain the SO nonlinear distortion field for the Pol-Mux system in time-domain as:

\begin{multline}
	\label{eqn47}
	\widetilde{u}_{2,x/y}(L,t)=\left(\frac{64}{81}\gamma^{2}P_{0}^{5/2}\sum_{m}\sum_{n}\sum_{k}\right.\\
	\left[\vphantom{\sum}2\left(a_{m,x/y}a_{m+n,x/y}^{*}\right.\right.
	\left.+a_{m,y/x}a_{m+n,y/x}^{*}\right)a_{n,x/y}\\
	\left(a_{k,x/y}a_{k,x/y}^{*}+a_{k,y/x}a_{k,y/x}^{*}\right)
	 \widetilde{g}_{m,n,k}^{\text{Term\,1}}(z,s,t)\\
	 +\left(a_{m,x/y}^{*}a_{m+n,x/y}+a_{m,y/x}^{*}a_{m+n,y/x}\right)a_{n,x/y}^{*}\\
	\times\left.\left.\left.\left(a_{k,x/y}a_{-k,x/y}+a_{k,y/x}a_{-k,y/x}\right)\widetilde{g}_{m,n,k}^{\text{Term\,2}}(z,s,t)\vphantom{\sum}\right]\vphantom{\frac{64}{81}\gamma^{2}P_{0}^{5/2}\sum_{m}\sum_{n}\sum_{k}}\right)\right|_{t=0}.
\end{multline}
After some simplifications, we obtain (\ref{eqn23}).



\ifCLASSOPTIONcaptionsoff
  \newpage
\fi

\end{document}